\documentclass[preprint,11pt,a4paper]{elsarticle}

\makeatletter
\@ifclassloaded{elsarticle}{}{
  \newcommand{\ead}[1]{}
  \newcommand{\address}[2][]{}
  \newtoks\@shromauth
  \newif\if@shromfirst \@shromfirsttrue
  \renewcommand{\author}[2][]{
    \if@shromfirst \@shromauth={#2}\@shromfirstfalse
    \else \@shromauth=\expandafter{\the\@shromauth\and #2}\fi}
  \def\@author@set#1{\gdef\@author{#1}}
  \newenvironment{frontmatter}{}
    {\expandafter\@author@set\expandafter{\the\@shromauth}\maketitle}
  \newenvironment{keyword}{\par\noindent\textbf{Keywords: }}{\par}
  \newcommand{\sep}{;\ }
  \usepackage[numbers]{natbib}
}
\makeatother

\usepackage{geometry}

\usepackage[utf8]{inputenc}
\usepackage[T1]{fontenc}
\usepackage{amsmath,amssymb,amsthm}
\usepackage{graphicx}
\usepackage{booktabs}
\usepackage{array}
\usepackage[table]{xcolor}
\usepackage{subcaption}
\usepackage[colorlinks=true,citecolor=blue,linkcolor=blue,urlcolor=blue]{hyperref}
\usepackage{cleveref}
\usepackage[expansion=false]{microtype}

\DeclareGraphicsExtensions{.pdf,.png,.jpg}

\usepackage{xspace}

\newcommand{\shrom}{SHROM\xspace}
\newcommand{\srvf}{\ensuremath{\mathrm{SRVF}}}
\newcommand{\latdim}{d}
\newcommand{\npts}{N}
\newcommand{\Rspace}{\mathbb{R}}
\newcommand{\qfun}{q}
\newcommand{\enc}{f_{\theta}}
\newcommand{\dec}{g_{\phi}}
\newcommand{\latent}{z}
\newcommand{\geodist}{d_{\mathcal{S}}}

\begin{document}

\begin{frontmatter}

\title{A shape-similarity latent space for fluid interfaces: invertible
reduced-order modelling of droplet morphology}

\author[cimne,upc1]{A.R.~Hashemi}
\ead{ahashemi@cimne.upc.edu}

\author[upc2]{M.R.~Hashemi}
\ead{mohammad.reza@upc.edu}

\author[cimne,upc1]{P.~Ryzhakov}
\ead{pavel.ryzhakov@upc.edu}

\address[cimne]{Centre Internacional de M\`etodes Num\`erics en
Enginyeria (CIMNE), C/Gran Capit\`a s/n, Campus Nord UPC, 08034
Barcelona, Spain}
\address[upc1]{Escola T\`ecnica Superior d'Enginyers de Camins, Canals i
Ports, Universitat Polit\`ecnica de Catalunya -- BarcelonaTech (UPC),
C/Jordi Girona 1, Campus Nord UPC, 08034 Barcelona, Spain}
\address[upc2]{Escola Polit\`ecnica Superior d'Enginyeria de Vilanova i
la Geltr\'u (EPSEVG), Fluid Mechanics Department, Universitat
Polit\`ecnica de Catalunya -- BarcelonaTech (UPC), Avinguda V\'ictor
Balaguer 1, 08800 Vilanova i la Geltr\'u, Spain}

\begin{abstract}

A droplet breaks up in tens of microseconds, and a high-speed recording
captures perhaps a dozen frames of it. The states in between cannot be
recovered without repeating the experiment, and simulating them is too costly
to sweep an operating envelope --- yet they are present in the data as a whole,
because a campaign spanning a device's actuation range produces morphologies
resembling those any single recording missed. Exploiting that requires a
representation which is low-dimensional, invertible, and faithful to shape
rather than to sampling. Proper orthogonal decomposition supplies the first two
but measures distance in sampled coordinates; manifold learning supplies the
third but no map back to a shape; elastic shape analysis supplies a shape
metric but no reduced coordinates.

We present \shrom{}, which composes all three. Interfaces are represented by
their square-root velocity functions, a neighbour graph is built over the
resulting shape space, and an autoencoder is trained to reconstruct while
penalising latents in which graph neighbours are not latent neighbours. The
demonstration uses $301{,}539$ inkjet droplet contours and four filmed break-up
sequences held out entirely.

The graph term does not improve reconstruction. It determines whether position
in the latent carries meaning: clustering the latent of an otherwise identical
model recovers the shape-space partition at chance
($\mathrm{ARI} = 0.063 \pm 0.059$), and at $0.781 \pm 0.049$ with the term
active. The trained model uses about two of its latent dimensions
($n_{\mathrm{eff}} = 1.20 \pm 0.02$) although the shape corpus is of higher
intrinsic dimension, and raising the latent to that dimension changes no
downstream result measurably. Waveform parameters predict the full closed
contour at $R^2 = 0.878 \pm 0.026$, and interpolation returns plausible
droplets at arbitrary intermediate times.

Negative results are reported in the same terms. The neighbour-graph metric
proved immaterial across five choices; interpolation error saturates at the
reconstruction limit, so a plain autoencoder leads that task by measuring the
fidelity the graph term trades for organisation. The principal limitation is
interpolation across a change of interface topology. No connected component of
the regularised latent contains both a single-component and a post-break-up
shape, at any regularisation strength or latent dimension tested, whereas an
unregularised autoencoder mixes them freely --- the separation is the
regulariser reporting a real discontinuity in shape space, not an artefact of
the embedding.

\end{abstract}


\end{frontmatter}

\section{Introduction}\label{sec:intro}

An inkjet droplet detaches, stretches into a ligament and breaks up in a few
tens of microseconds. An engineer choosing an actuation waveform needs to know
what it will produce, and an engineer diagnosing a failed print needs to know
what happened between two frames of a high-speed recording. Neither question is
easy to answer. Simulation resolves the intermediate states but at a cost per
configuration that rules out searching an operating envelope. Imaging records
them directly, but every campaign fixes a frame rate, and the states between
consecutive frames of an existing recording are not available at any price ---
they would require the experiment to be repeated.

The states are missing from any single recording. They are not, however,
missing from the data as a whole. A campaign that sweeps the achievable range
of actuation parameters produces hundreds of thousands of droplets, and a
morphology one droplet passes through unobserved closely resembles one that a
different setting produced and the camera did capture. \emph{The observations
are sparse in time and dense in shape.} If a model could tell which shapes
resemble which, it could supply from the corpus what any one sequence failed to
record --- and answer the same question in reverse, predicting the shape a
waveform will produce because similar waveforms produced similar shapes.

That is what this paper builds, and it is why a representation is needed rather
than a lookup. Such a representation must be \emph{low-dimensional}, or there
is nothing compact to interpolate through or regress onto. It must be
\emph{invertible}, so that a coordinate can be turned back into an interface
rather than merely scored. And it must be \emph{geometry-faithful}, so that
proximity in the representation means similarity of shape rather than
similarity of sampling --- because resemblance is the entire mechanism by which
a missing state is recovered.

No established tool supplies all three. Reduced-order models built on
proper orthogonal decomposition \citep{maulik2021reducedorder} are invertible
and compact but measure distance between sampled coordinates, where a curve and
a reparametrisation of it appear far apart. Manifold learning
\citep{mcinnes2018umap} respects an arbitrary input metric but provides no map
back to a shape. Elastic shape analysis \citep{srivastava2011srvf} supplies a
genuine shape metric with a full statistical apparatus, but is a representation
rather than a reduction, and yields no low-dimensional coordinates. \Cref{sec:background}
develops each of these and what it lacks.

\subsection*{Approach}

This work composes the three into one model. Interfaces are represented by
their square-root velocity functions, so that distance between shapes is a
geodesic on a sphere rather than a norm on a coordinate vector. A neighbour
graph is built over that shape space. An autoencoder is then trained with a
reconstruction term and a term that penalises the latent when graph neighbours
are not latent neighbours. This supplies the inverse map that manifold learning
lacks, together with the shape geometry that a plain autoencoder does not
acquire.

Every ingredient is established, and \Cref{sec:background} cites the precedent
for each. The composition achieves what none of the parts achieves alone. Building the
neighbour graph over an elastic shape space produces a latent in which
\emph{position} corresponds to morphology, so that a coordinate between two
observations decodes to the shape belonging at that coordinate. This is
what converts density in shape space into predictive power along a sparsely
sampled trajectory, and it is invisible to reconstruction error --- a decoder
can return every training shape faithfully while the space between those shapes
holds nothing.

That claim is testable and we test it directly, against an otherwise identical
model with the graph term removed, and against a baseline that pursues latent
geometry through a different mechanism.

\subsection*{Demonstration}

The method is demonstrated on drop-on-demand inkjet printing, which supplies
the conditions the argument requires. The morphology family is rich enough that
resemblance is informative, spanning compact droplets, long ligaments and
post-break-up satellites. A natural eight-parameter actuation input covers the
device's achievable range. The filmed sequences through break-up are sparse in
time, and can therefore be used to test whether the missing states are
recoverable. Our own earlier work on this system
predicted droplet descriptors and accompanying-structure classes from actuation
parameters \citep{Hashemi_2025,Ares_2025}; the present model predicts the full
closed interface geometry, which those approaches do not attempt.

The demonstration is a demonstration. Nothing in the method refers to droplets,
and \Cref{sec:disc:generality} states plainly which properties a different
domain would need for it to transfer, and that such transfer is untested here.

\subsection*{Contributions}

\begin{itemize}
  \item An elastic (\srvf{}) shape representation of a fluid interface used as
        a learning target. The \srvf{} has been used this way before, for
        protein backbones \citep{huang2021gvae}; elastic shape analysis is
        mature for biological and medical shapes generally. We are not aware of
        prior use on fluid interfaces, where data-driven shape work has instead
        used Fourier descriptors or raw images.

  \item A neighbour graph built over that shape space, rather than over raw or
        Fourier coordinates, as the regularisation signal. The claim
        concerns the space, not the distance function: an ablation across five
        graph metrics finds no difference, so what matters is that proximity
        means shape similarity, not which formula computes it.

  \item An invertibility and organisation assessment rather than an assertion:
        the graph term is ablated against an identical model, the latent is
        probed by decoding coordinates never seen in training, and the
        reported diagnostics are defined before they are measured.

  \item Three downstream demonstrations, reported with their null results:
        interpolation through break-up, prediction of geometry from actuation
        parameters, and recovery of morphological structure. On the first, a
        plain autoencoder performs better. The cause is identified rather than
        left open: that task's residual error is reconstruction error, which the
        manifold term trades for latent organisation. Establishing that a proof
        of concept measures a quantity other than the one it was designed to
        test is itself a result.
\end{itemize}

We are explicit about what is not claimed. This is not the first autoencoder
for droplet shapes, the first graph-regularised latent, or a new interpolation
scheme; \Cref{sec:bg:positioning} attributes each to its precedent.

\subsection*{Structure}

\Cref{sec:background} reviews the four candidate approaches and identifies what
each lacks. \Cref{sec:method} defines the method in domain-independent terms,
and \Cref{sec:setup} instantiates it, giving every parameter a value and a
reason. \Cref{sec:results} reports the ablation first, since it is what
justifies the method, followed by the three demonstrations, the baseline
comparison and two configuration studies. \Cref{sec:discussion} synthesises,
including the central limitation --- interpolation across a change of interface
topology --- and \Cref{sec:conclusions} states what was and was not
established. \Cref{sec:appendix} holds the supporting measurements: the latent
connectivity that underlies the topology argument, and a set of latent-geometry
diagnostics that turned out not to discriminate between configurations, kept
because the null is worth recording.

\section{Theoretical background}\label{sec:background}

Recovering an unobserved interface state from a corpus of observed ones
requires four things at once: a representation in which distance between shapes
is meaningful, a low-dimensional coordinate system, an inverse from those
coordinates back to a shape, and a latent whose \emph{organisation} reflects
shape similarity. The first three are familiar requirements; the fourth is the
one that makes resemblance usable, and it is the one no established method
supplies.

This section takes the candidates in turn and asks of each the same question:
given two observed droplets and a state between them that was never recorded,
what stops this method from supplying it? The method of \Cref{sec:method} is
assembled from precisely these pieces and claims novelty only in their
composition.

\subsection{Elastic shape analysis and the \srvf{}}\label{sec:bg:srvf}

Comparing two curves requires deciding when they are the same shape. Two
outlines that differ only by translation, scale, rotation, or by where the
parametrisation begins and how quickly it traverses the boundary, are
geometrically identical but numerically far apart. Elastic shape analysis
removes these by construction \citep{srivastava2011srvf,srivastava2016functional}.

The square-root velocity function of \Cref{eq:srvf} has the property that
reparametrisation acts on it as an isometry of $\mathbb{L}^2$, so the quotient
by reparametrisation is well behaved rather than requiring a penalty term.
After normalisation, shapes are points on a unit Hilbert sphere and the
distance between them is the great-circle angle. The framework carries a
developed statistical apparatus --- means, geodesics, tangent-space principal
component analysis, hypothesis tests --- and has been applied to proteins,
neuronal trees, arteries and biological outlines
\citep{kurtek2012statistical,duncan2018neuronaltrees,guo2022brainarterial}.

Why this alone is insufficient. The \srvf{} is a representation, not a
reduced model. It replaces one high-dimensional description of a curve with
another of the same dimension, in which distances are meaningful. It can tell
us that two droplets resemble one another; it cannot produce the state between
them, because there are no coordinates to move along. Everything that follows exists to obtain those
coordinates without giving up the metric.

\subsection{Dimensionality reduction and the invertibility gap}\label{sec:bg:dr}

The obvious next step is a dimensionality reduction, and the two families
available fail in complementary ways.

Linear methods --- proper orthogonal decomposition and its equivalents --- are
the standard tool for reduced-order modelling of fluid systems
\citep{maulik2021reducedorder,fresca2022pod} and are invertible by
construction. But they operate on sampled coordinates, and Euclidean distance
between sampled coordinates is exactly the quantity elastic shape analysis
exists to avoid: a curve and its reparametrisation are far apart in that
metric, so the decomposition spends components describing parametrisation
rather than shape.

Nonlinear manifold learning --- Isomap, locally linear embedding, $t$-SNE and
UMAP \citep{tenenbaum2000global,roweis2000nonlinear,mcinnes2018umap} --- can
respect an arbitrary input metric, including a geodesic one, and produces
embeddings that capture nonlinear structure well. The obstacle is invertibility. These methods assign coordinates to the points
they were given and provide no map back: a latent coordinate cannot be turned
into a curve. They can place the missing state among its neighbours and cannot
say what it looks like.
For visualisation this is immaterial. For a reduced-order model, whose entire
purpose is to be evaluated at coordinates that were never observed, it is
disqualifying.

\subsection{Autoencoders and the geometry gap}\label{sec:bg:ae}

An autoencoder closes the invertibility gap by construction: the decoder is the
inverse map, trained jointly with the encoder \citep{goodfellow2016deep}.
Autoencoders are now routine in non-intrusive reduced-order modelling
\citep{kehls2025nonintrusive,san2018artificial} and have been applied directly
to droplet outlines, both as convolutional models over boundary images
\citep{khor2019emulsion} and over Fourier coefficients
\citep{durve2024fourier}.

Why this alone is insufficient. A reconstruction objective constrains the
composition of encoder and decoder at the training points and says nothing
about the space between them. A model can reconstruct its data faithfully while
arranging the latent arbitrarily --- and, in particular, while mapping the
regions between clusters of training points to shapes that are not plausible
members of the family. Reconstruction error cannot detect this, because it is
never evaluated there.

This is precisely the failure that matters here. The state between two captured
frames lies, by construction, in the space \emph{between} observations. An
autoencoder trained to return its inputs faithfully has been given no reason to
put anything sensible there, and no reason to place a droplet's unobserved
morphology near the similar droplets that were observed. Latent position need
not carry meaning, and without it the corpus's density in shape space cannot be
brought to bear on any one sequence's sparsity in time.
\Cref{sec:res:ablation} measures the size of this effect directly.

\subsection{Graph regularisation of a latent space}\label{sec:bg:graph}

The response is to add a term that constrains latent organisation explicitly. A
neighbour graph is built over the data, and the latent is penalised when
neighbours in that graph are not neighbours in the latent. Parametric UMAP
\citep{sainburg2021parametric} learns a parametric encoder against the UMAP
objective; topological autoencoders \citep{moor2020topological} preserve
persistent-homology features; geometry-regularised autoencoders
\citep{duque2022geometry} regularise the bottleneck with diffusion-potential
distances; and isometry-regularised variants
\citep{atzmon2020isometric,kato2020rate} constrain the decoder Jacobian instead.

This category is not new, and we make no claim on it. The
architectural idea of regularising an autoencoder latent with a
neighbourhood-derived term is established, and several of the works above
predate this one by years. What differs here is the space the graph is built
over: the neighbours are neighbours in an elastic shape space rather than in
pixel, Fourier or raw-coordinate space, so the structure the regulariser
imposes is shape structure. That is the mechanism by which one droplet's
recorded morphology can stand in for another's unrecorded one --- the graph is
what makes resemblance operational rather than merely measurable.

A caution on how narrowly to state even that. Our own ablation
(\Cref{sec:res:metric}) shows that the specific distance function used to build
the graph is immaterial across five choices --- including choices that produce
demonstrably different graphs. The defensible claim is therefore about the
\emph{space}, not the metric: what matters is that proximity in the graph means
shape similarity, not which formula computes it.

\subsection{Multi-component shapes}\label{sec:bg:multi}

A closed-curve \srvf{} represents one closed curve. An interface undergoing
breakup does not remain one, and the moment of breakup is usually the
phenomenon of interest, so this is not an edge case that can be excluded.

Four approaches exist. \emph{Elastic shape graphs}
\citep{guo2020elasticgraphs,basubal2022shapegraphs} extend the elastic
framework to graph-structured objects with nodes and edges, and are the most
principled available answer. The \emph{extended \srvf{}}
\citep{wang2023esrvf} admits branch birth and death, developed for
three-dimensional tree structures such as neurons and vasculature.
\emph{Currents and varifolds} \citep{charon2013varifold,hsieh2021varifoldquantization}
represent shapes as measures, which accommodates topology change naturally.
\emph{Unbalanced optimal transport} in the Wasserstein--Fisher--Rao geometry
allows mass to be created and destroyed between configurations.

A difficulty common to all four, and to the route taken here, is that a latent
space is usually Euclidean while a shape space with breakup events is not.
\citet{syrota2024decoder} identify exactly this mismatch --- a compact data
manifold embedded in a Euclidean latent --- as a limitation of pullback-metric
geodesics, which is the class of method \Cref{sec:res:poc1} evaluates. The
mismatch is not created by any particular construction; it is a property of
asking one connected coordinate system to hold shapes that are genuinely
disconnected.

This work takes none of the four. It joins components with a degenerate
connector (\Cref{sec:method:bridge}), which is the simplest construction that
returns a single closed curve and keeps the entire \srvf{} apparatus intact.
This is a modelling choice with a measurable cost, and it is stated
here so that the limitation reported in \Cref{sec:disc:limitation} arrives
prepared rather than as a late concession. The reasons for setting the
alternatives aside --- implementation availability, dimensional mismatch, and
loss of the reparametrisation guarantees the results depend on --- are given in
\Cref{sec:disc:alternatives}.

\subsection{Positioning}\label{sec:bg:positioning}

Every component of the method described in \Cref{sec:method} is prior art. The \srvf{} is two decades old, and has already been used as an autoencoder
target. \citet{huang2021gvae} represent proteins as \srvf{}s of
three-dimensional open curves and encode them onto a low-dimensional latent
hypersphere, using a von Mises--Fisher prior because a Gaussian one mismatches
the spherical geometry. They decode back to structures, and observe that
geodesics on shape space can be approximated by interpolation in the latent. The interpolation thesis of
this paper is therefore not new in itself; what differs is the domain --- planar
fluid interfaces rather than protein backbones --- and the mechanism. A neighbour graph is a dataset-wide constraint specifying which shapes must
remain close together, whereas a distributional prior constrains only the shape
of the latent density. \citet{huang2021gvae} also run no ablation separating
latent organisation from reconstruction, which is the measurement
\Cref{sec:res:ablation} reports. Autoencoders for reduced-order modelling are
routine, including for droplet outlines specifically. Graph-regularised latents
are an established category. Interpolating along a pullback geodesic is a known
technique. Predicting a reduced representation from process parameters, and
clustering in a learned latent, have both been done for droplets before
\citep{durve2024fourier}.

The contribution is the composition, together with evidence that it
buys something measurable. Specifically, building the neighbour graph
over an elastic shape space, rather than over raw or Fourier coordinates,
produces a latent in which position corresponds to morphology. Reconstruction
error does not measure this property, so no autoencoder acquires it by
default. \Cref{sec:res:ablation} isolates it against an otherwise
identical model, and \Cref{sec:res:benchmark} against a baseline that optimises
a different notion of latent geometry.

We are also careful about what is \emph{not} claimed. This is not the first
autoencoder for droplet shapes, nor the first use of a neighbourhood term to
regularise a latent, nor a new interpolation scheme. Where a proof of concept
has a close precedent, it is cited at the point of use rather than in a
contributions list.

\section{Methodology}\label{sec:method}

This section specifies the method independently of any dataset. It defines the
shape representation, the neighbour graph built over it, the regularised
training objective, and the diagnostics and evaluation protocols used later.
Every parameter introduced here is given a value in \Cref{sec:setup}.

\subsection{Overview}\label{sec:method:overview}

\shrom{} turns a corpus of observed interfaces into a space that can be
navigated: every coordinate in it decodes to an interface, and coordinates near
one another decode to interfaces that resemble one another. A state that no
recording captured is then obtainable by moving to where it belongs.

\begin{figure}[t]
  \centering
  \includegraphics[width=\textwidth]{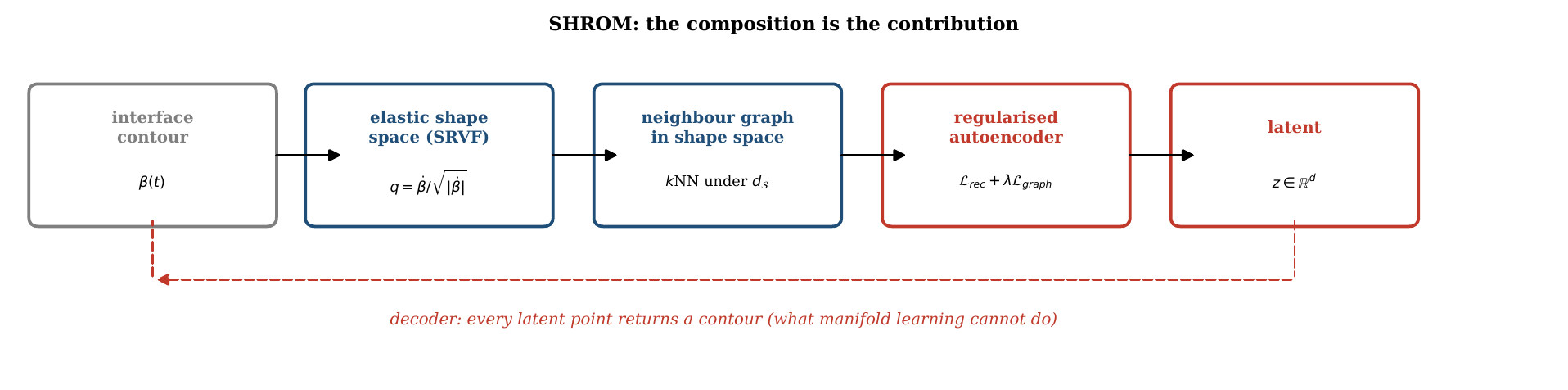}
  \caption{The \shrom{} pipeline. The forward path composes an elastic shape
  metric, a neighbour graph built over that space, and an autoencoder. The
  dashed return path is the decoder: every latent point maps back to a contour,
  which is what distinguishes this from manifold learning.}
  \label{fig:pipeline}
\end{figure}

Three components produce this, each individually standard and, to our
knowledge, not previously combined. These are an elastic shape metric, a
neighbour graph built over the resulting shape space, and an autoencoder whose
loss is regularised by that graph. The forward path takes a closed planar interface to
a low-dimensional code,
\begin{equation}\label{eq:pipeline}
  \beta \;\longrightarrow\; \qfun \;\longrightarrow\;
  \mathcal{G} \;\longrightarrow\; \latent = \enc(\qfun),
\end{equation}
and the decoder $\dec$ returns every latent point to a contour. That return
path is what separates this from manifold learning: an embedding assigns
coordinates to the data it was given, whereas a reduced-order model must also
answer what lies at a coordinate it has never seen. \Cref{fig:pipeline} shows
the composition.

The design commits to one idea: that the latent should be organised by shape
similarity, and that reconstruction error alone will not produce such an
organisation. A decoder can reconstruct its training data faithfully while
mapping the space between those points to nothing meaningful --- and it is
exactly that space which holds the states no camera recorded. The graph term
exists to make \emph{position} in the latent carry information, and
\Cref{sec:res:ablation} measures what that buys.

\subsection{Shape representation}\label{sec:method:srvf}

A closed planar interface is a curve $\beta : [0,1] \to \Rspace^2$. Working
with $\beta$ directly is unsatisfactory for statistics, because two curves that
differ only in where the parametrisation starts, or in how fast it traverses
the outline, are the same shape but different functions. The square-root
velocity function removes the latter,
\begin{equation}\label{eq:srvf}
  \qfun(t) \;=\; \frac{\dot\beta(t)}{\sqrt{\lVert \dot\beta(t) \rVert}},
\end{equation}
under which a reparametrisation of $\beta$ acts on $\qfun$ by an isometry of
$\mathbb{L}^2$ \citep{srivastava2011srvf}. Translation is removed by centring
on the area-weighted centroid, and scale by dividing out the curve length. What
remains is normalised to unit $\mathbb{L}^2$ norm, so every shape is a point on
the unit Hilbert sphere $\mathcal{S}$, where the distance between two shapes is
the great-circle angle
\begin{equation}\label{eq:geodesic}
  \geodist(\qfun_1, \qfun_2)
    \;=\; \arccos \langle \qfun_1, \qfun_2 \rangle .
\end{equation}
This is the property the rest of the method rests on: distance between shapes
is a geometric quantity on a sphere, not an arbitrary norm on a coordinate
vector.

Two consequences follow.

First, normalisation is not lossy. The discarded scale factor is
retained alongside each code, so a contour is recoverable at its original size;
the sphere carries shape, and scale travels beside it as a scalar.

Second, \Cref{eq:geodesic} is an upper bound on the true elastic
distance. The elastic metric is defined as an infimum over rotations and
reparametrisations, and we optimise over neither: the seam is fixed by a
deterministic rule (\Cref{sec:method:bridge}) and rotation is left as observed.
What \Cref{eq:geodesic} computes is therefore the \emph{unaligned} distance
$d_0 \geq d$. We use it deliberately --- it is deterministic, cheap, and
comparable across a large dataset, where an alignment search is neither ---
but the gap is a genuine approximation and worth
quantifying. Optimising jointly over cyclic shift and rotation reduces the
measured distance by $19.7\%$ on average, and allowing a bounded elastic warp
in addition by $32.4\%$. The gap is smaller between similar shapes
($14.6\%$ within a topology class) than between dissimilar ones ($22.2\%$
across classes), which is the expected behaviour, but it is material in both
regimes. Distances reported in this paper are therefore consistently
\emph{over}-estimates, by a factor that is stable enough not to reorder
comparisons.

\subsection{Assembling multi-component interfaces}\label{sec:method:bridge}

An interface undergoing breakup is not one closed curve. After pinch-off the
outline consists of two or more disjoint components, and the \srvf{}
representation of \Cref{sec:method:srvf} is defined for a single closed curve
only.

We assemble the components into one closed curve by joining them with a
\emph{degenerate connector}: each component is rolled so that its seam vertex
faces the other, and the components are then traversed as a single closed
path. The connector is a segment of zero enclosed width, traversed once in each
direction, so it contributes no area. Each component is resampled to an equal
share of the point budget before concatenation, so that both bodies carry equal
parametrisation weight regardless of their relative size --- without this, a
small satellite would occupy a negligible fraction of the curve and be
effectively invisible to the metric.

The connector is a modelling choice, not an established technique. We
are not aware of a precedent for representing a multi-component interface this
way within an elastic metric, and it is not claimed to be the principled
solution. Its cost is measurable: the connector places shapes of differing
topology far apart in the resulting space, and \Cref{sec:disc:limitation}
quantifies that separation. Interpolation \emph{across} a change of topology is
correspondingly unreliable, and is the method's principal limitation.

The construction requires the two seam vertices to remain coincident after
assembly, which constrains the differencing scheme. Velocities are computed by
forward differences, assigning to each vertex its outgoing step; a centred
difference displaces every vertex by half a segment, which separates the
coincident seam vertices and reopens the connector into two disjoint legs.

\subsection{The shape-space neighbour graph}\label{sec:method:graph}

The regulariser needs a notion of which shapes are neighbours. We build a
$k$-nearest-neighbour graph over the \srvf{} coordinates and convert it to a
fuzzy simplicial set in the manner of UMAP \citep{mcinnes2018umap}, giving a
weighted graph $\mathcal{G}$ whose edge weights express membership strength
rather than distance.

The graph is built with Euclidean distance on \srvf{} coordinates,
not with \Cref{eq:geodesic}. What makes $\mathcal{G}$ a \emph{shape-space}
graph is the space it is computed over rather than the distance function: the
coordinates are \srvf{}s, so its neighbours are neighbours in shape space
rather than in pixel or Fourier space.

The two choices coincide. Because \srvf{}s are unit-norm, the
chord and the arc are related by
\begin{equation}\label{eq:chord-arc}
  \lVert \qfun_1 - \qfun_2 \rVert
    \;=\; 2 \sin\!\left( \tfrac{1}{2}
      \geodist(\qfun_1, \qfun_2) \right),
\end{equation}
which is strictly increasing in $\geodist$ on $[0,\pi]$. A strictly monotone
transform cannot reorder anything, so the Euclidean $k$ nearest neighbours
\emph{are} the geodesic $k$ nearest neighbours --- the graph is the geodesic's
own, obtained without evaluating \Cref{eq:geodesic}. \Cref{sec:res:metric}
verifies this and further shows that the downstream results are insensitive to
the choice of graph metric altogether.

\subsection{The regularised objective}\label{sec:method:loss}

The encoder $\enc$ and decoder $\dec$ are trained against
\begin{equation}\label{eq:loss}
  \mathcal{L}
    \;=\; \mathcal{L}_{\mathrm{recon}}
      \;+\; \lambda \, \mathcal{L}_{\mathcal{G}},
\end{equation}
where $\mathcal{L}_{\mathrm{recon}}$ is the mean squared error between input
and reconstructed \srvf{}, and $\mathcal{L}_{\mathcal{G}}$ is a graph term
evaluated on sampled edges.

The graph term is the UMAP cross-entropy. For a latent pair $(\latent_i,
\latent_j)$ define the latent-space affinity
\begin{equation}\label{eq:qij}
  q_{ij} \;=\; \bigl( 1 + a \lVert \latent_i - \latent_j \rVert^{2b}
    \bigr)^{-1},
\end{equation}
and let $\mathcal{L}_{\mathcal{G}}$ be
\begin{equation}\label{eq:umaploss}
  \mathcal{L}_{\mathcal{G}}
    \;=\; -\!\!\!\underset{(i,j)\,\sim\,\mathcal{G}}{\mathbb{E}}\!\!
        \log q_{ij}
      \;-\; \nu \!\!\underset{(i,j)\,\sim\,\mathrm{neg}}{\mathbb{E}}\!\!
        \log \bigl( 1 - q_{ij} \bigr),
\end{equation}
an attraction over edges drawn from $\mathcal{G}$ in proportion to their
weight, and a repulsion over uniformly sampled non-edges with strength $\nu$.
Sampling edges by weight lets the attraction term omit an explicit weight
factor: strong edges are simply drawn more often. Both expectations are
averaged rather than summed, so that $\lambda$ retains its meaning when the
edge-batch size changes.

The kernel parameters $a$ and $b$ are not free. They are fitted so that
\Cref{eq:qij} approximates a piecewise curve that is flat out to a chosen
separation $\texttt{min\_dist}$ and decays with width $\texttt{spread}$
thereafter; $\texttt{min\_dist}$ is thus the interpretable knob and $(a,b)$
follow from it. Its effect is a locality--globality trade-off ---
small values pack neighbours tightly and separate groups, large values
prioritise a connected embedding --- which \Cref{sec:res:umap} examines.

Finally, $\lambda$ is annealed rather than held fixed. Training begins with the
graph term effectively inactive and raises $\lambda$ through a sequence of
phases. The reason is ordering: a randomly initialised decoder produces
meaningless reconstructions, so a graph term applied from the first step
organises a latent that does not yet correspond to anything. Letting
reconstruction establish itself first, then imposing structure, avoids
optimising the two objectives against each other while both are uninformed.

\subsection{Decoding and invertibility}\label{sec:method:decode}

The decoder maps a latent point to a $2\npts$-vector, which is renormalised to
unit norm, so the output lies on $\mathcal{S}$ by construction. A contour is recovered from an \srvf{} by inverting
\Cref{eq:srvf} --- multiplying each velocity by its own magnitude and
integrating,
\begin{equation}\label{eq:invert}
  \dot\beta(t) = \qfun(t)\,\lVert \qfun(t) \rVert,
  \qquad
  \beta(t) = \beta(0) + \int_0^t \dot\beta(s)\,\mathrm{d}s,
\end{equation}
discretised as a cumulative sum. Restoring the stored scale factor returns the
contour at its original size. Every latent point therefore yields a contour;
whether it yields a \emph{plausible} one is an empirical question, and
\Cref{sec:res:cleaning} answers it by decoding a regular lattice.

\subsection{Diagnostics}\label{sec:method:diagnostics}

Four quantities are used to characterise a trained latent. All are defined here
and measured in \Cref{sec:results}.

\paragraph{Effective dimensionality.}
The participation ratio of the \emph{squared} decoder-Jacobian singular values,
$n_{\mathrm{eff}} = \bigl(\sum_i \sigma_i^2\bigr)^2 / \sum_i \sigma_i^4$,
which counts how many latent directions the decoder actually uses. Squaring
weights the ratio by the variance each direction carries rather than by its
amplitude. A latent of nominal dimension
$\latdim$ may have $n_{\mathrm{eff}}$ well below $\latdim$.

\paragraph{Isometry defect.}
The deviation of $J^\top J$ from a scaled identity, measuring how anisotropically
the decoder stretches latent space. It is reported for its own sake and because
it is the quantity an isometry-regularised baseline optimises directly
(\Cref{sec:setup:baselines}).

\paragraph{$k$NN structure overlap.}
The fraction of each shape's $k$ nearest neighbours in shape space that remain
among its $k$ nearest in the latent. This is the direct measure of whether the
latent preserves shape neighbourhoods, and the axis on which we claim an
advantage.

\paragraph{Trustworthiness.}
The standard rank-based penalty for points that are close in the latent but far
in shape space \citep{zhang2021pydrmetrics}. It is reported alongside structure
overlap rather than instead of it, since a good score on either alone does not
imply a useful embedding \citep{kazempour2024good}.

\subsection{Reading the reported quantities}\label{sec:method:scales}

Every metric used in this paper is dimensionless, and several are unfamiliar
outside their own literature. This subsection states what each one measures,
its range, and what value would count as trivial, so that the numbers in
\Cref{sec:results} can be read without reference to their source papers.

\begin{table}[t]
  \centering
  \caption{The reported quantities, their ranges, and their reference points.
  ``Chance'' is the value a model carrying no information would obtain.}
  \label{tab:scales}
  \begin{tabular}{lll>{\raggedright\arraybackslash}p{0.34\textwidth}}
    \toprule
    Quantity & Range & Chance / trivial & Meaning \\
    \midrule
    Geodesic distance $\geodist$ & $[0, \pi]$ rad & --- &
      Angle between two shapes on the pre-shape sphere. Zero is identical. \\
    Adjusted Rand index & $[-1, 1]$ & $0$ &
      Agreement between two partitions, corrected for chance. \\
    $k$NN structure overlap & $[0, 1]$ & $k/n \approx 0$ &
      Fraction of a shape's $k$ nearest neighbours preserved in the latent. \\
    Trustworthiness & $[0, 1]$ & $\approx 0.5$ &
      Penalty for latent neighbours that are distant in shape space. \\
    Silhouette & $[-1, 1]$ & $0$ &
      Cluster separation relative to within-cluster spread. \\
    Effective dimension $n_{\mathrm{eff}}$ & $[1, \latdim]$ & $\latdim$ &
      Latent directions the decoder actually uses. \\
    Solidity & $(0, 1]$ & $1$ (convex) &
      Enclosed area divided by convex-hull area. \\
    \bottomrule
  \end{tabular}
\end{table}

Two conventions apply throughout. Reconstruction error is mean squared error on
unit-norm \srvf{} vectors, so it is small in absolute terms by construction and
is meaningful only as a ratio between models. And $\geodist$ is quoted in
radians, for which the corpus itself supplies the natural scale: two shapes
drawn at random from the dataset are separated by $1.06 \pm 0.55$~rad
(median $1.04$). A geodesic error of $0.1$~rad is therefore about a tenth of
the distance between two arbitrary droplets, and errors are given alongside
this reference wherever they first appear.

\subsection{Evaluation protocol}\label{sec:method:protocol}

\paragraph{Interpolation.}
Two protocols, answering different questions. In the \emph{held-out anchor}
protocol, a path is fitted through a sparse subset of a time series and
evaluated at the withheld frames, testing whether the latent supports
reconstruction of unobserved states. In the \emph{leave-one-out} protocol, a
single frame is withheld and a path fitted through all remaining frames,
isolating the error at one point from the sparsity of the fit. Paths are
compared against a great-circle reference on $\mathcal{S}$, which is the
geometric baseline the representation itself suggests --- not a claim about
physical evolution.

In the \emph{adjacent-pair} protocol, two consecutive frames are joined by a
straight line between their latent codes. Through two points a straight line is
the exact interpolant, so this protocol has no fitting freedom at all: no
spline, no anchor spacing, nothing to tune. It therefore separates what the
representation costs from what the path fit costs, which the other two
protocols confound.

Averaging latent codes is legitimate in a way that averaging contours is not,
and the distinction is worth stating. The pointwise mean of two contours is not
in general a valid interface: it depends on how each curve was sampled, and it
lies off the shape space rather than on it. The correct notion of a mean on
$\mathcal{S}$ is the Fr\'echet (Karcher) mean, which minimises the summed
squared geodesic distance and must be found iteratively. The latent, by
contrast, is an ordinary Euclidean space, so the midpoint of two codes is
well defined and cheap. The decoder returns from that midpoint a valid shape lying near both
endpoints, which is what the interpolation tasks require. That shape is not the
Fr\'echet mean of the two endpoints. The decoder is nonlinear, so it does not carry Euclidean
midpoints to geodesic midpoints.

\paragraph{Shape error.}
Three metrics, because each is blind to something. The geodesic
(\Cref{eq:geodesic}) is the metric the model is trained under and the primary
figure. The Hausdorff distance reports the single worst point disagreement, so
a small local defect that the geodesic averages away is visible. The Chamfer
distance averages nearest-point distances and is therefore forgiving of
outliers but sensitive to systematic offsets. A prediction that is good on all
three is good; disagreement among them is diagnostic, and we report all three
throughout.

\section{Experimental setup}\label{sec:setup}

The argument of \Cref{sec:intro} rests on a specific condition: that the corpus
is dense in shape even where any one recording is sparse in time. This section
instantiates the method on a dataset built to satisfy that condition, and gives
every parameter of \Cref{sec:method} a value and a reason.

\subsection{Data}\label{sec:setup:data}

The interfaces are drop-on-demand inkjet droplets recorded by high-speed
imaging. Each frame is a single actuation captured at a fixed delay, so the corpus
samples the droplet morphology family broadly rather than following any one
droplet in time. This is the design that makes the corpus dense in shape. A morphology that one
droplet passes through between two frames of its own sequence is, with high
probability, close to a morphology produced and captured at some other actuation
setting. Every frame carries the eight waveform parameters that
produced it --- three voltages $V_1, V_2, V_3$, three pulse widths $w_1, w_2,
w_3$ and two dwell times $d_1, d_2$ --- which are the inputs for the
waveform-to-shape task of \Cref{sec:res:poc2}.

Of the $316{,}335$ extracted frames, $44.5\%$ contain more than one component:
a droplet that has already broken up, or is in the act of doing so. These are
not defects to be discarded but the physically interesting cases, and they are
retained through the connector of \Cref{sec:method:bridge}.

Four further sequences, each of roughly $1{,}000$ consecutive frames, follow
individual droplets through time. These form the time-series set used for the
interpolation task of \Cref{sec:res:poc1}. They are a separate acquisition and
appear in no training set; they are held out entirely.

\subsection{Two-stage cleaning}\label{sec:setup:cleaning}

Automated contour extraction produces some outlines that are not droplets, and
some that are droplets but are unusable. Screening happens in two stages
because the two kinds of defect become visible at different points: stage~1
judges each raw component \emph{before} assembly, while stage~2 judges the
assembled shape, and some defects exist only after assembly.

\paragraph{Stage 1: per-component screening.}
Five criteria, applied to each component as extracted. Together they reject
$11{,}747$ frames, $3.71\%$ of the corpus, leaving $304{,}588$
(\Cref{tab:cleaning}).

\begin{table}[t]
  \centering
  \caption{Stage-1 rejection categories. $11{,}747$ of $316{,}335$ frames
  rejected ($3.71\%$), leaving $304{,}588$.}
  \label{tab:cleaning}
  \begin{tabular}{lrp{0.42\textwidth}}
    \toprule
    Criterion & Rejected & Rationale \\
    \midrule
    \texttt{branched\_tail}   & 10{,}110 & The tail doubles back on itself and encloses no area: an extraction artefact, not a ligament. \\
    \texttt{area\_too\_small} & 612      & Below a lower area cut separated from the bulk of the distribution by a genuine gap. \\
    \texttt{all\_components\_defective} & 528 & Every component of the frame failed one of the other criteria, leaving nothing to assemble. \\
    \texttt{frame\_edge\_crop} & 287     & Part of the outline is the image border rather than the interface. \\
    \texttt{bare\_filament}   & 210      & A ligament with no droplet body attached. \\
    \bottomrule
  \end{tabular}
\end{table}

\begin{figure}[t]
  \centering
  \includegraphics[width=0.85\textwidth]{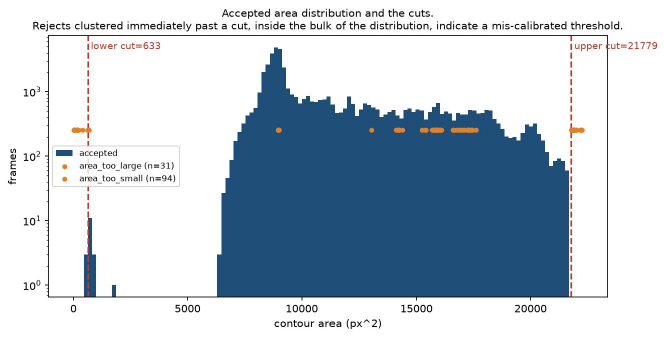}
  \caption{Accepted area distribution with the cuts drawn. The lower cut
  separates a genuinely distinct population; the upper cut sat inside the
  continuous bulk and was removed.}
  \label{fig:rejections}
\end{figure}

One further criterion, an \emph{upper} area cut, was applied in an earlier
configuration and has been removed. Its threshold sat inside the continuous
bulk of the area distribution rather than beyond a gap, and inspection of the
shapes it rejected found them to be healthy long-tailed droplets --- among the
most physically interesting in the corpus. \Cref{fig:rejections} shows the
accepted area distribution with both cuts drawn: the lower cut separates a
genuinely distinct population, the upper one did not.

\paragraph{Stage 2: per-cluster outlier removal.}
Some shapes survive stage~1 as valid components yet assemble into an
implausible whole. These are not identifiable by any per-component test, so
stage~2 works in shape space: the corpus is clustered, the geodesic distance
from each shape to its cluster mean is computed, and shapes beyond a
percentile of that within-cluster radius are removed. At the $99.0$th
percentile this removes $1.00\%$ per cluster, $3{,}049$ shapes, leaving
$\mathbf{301{,}539}$.

The percentile is applied \emph{per cluster} rather than globally because the
clusters differ in spread; a single global radius would over-cut the compact
clusters and under-cut the diffuse ones. A fixed percentile nonetheless removes
the same fraction from every cluster regardless of how many defects it actually
contains. A per-cluster adaptive threshold would be preferable and is left to
future work.

Two criteria are deliberately \emph{absent}, and both were tried and rejected.
Size is not a stage-2 criterion, for the reason above. Self-intersection is not
a criterion either: a shape assembled through the connector of
\Cref{sec:method:bridge} is self-intersecting by construction, so the test
flags the representation rather than the shape.

\subsection{Representation and architecture}\label{sec:setup:arch}

Each contour is resampled to $\npts = 160$ points, giving a $320$-dimensional
\srvf{} vector; velocities use the forward differencing of
\Cref{sec:method:bridge}. Multi-component frames are assembled by the
\texttt{legacy} policy: the two largest components are ordered left-to-right,
each is resampled to $\npts/2$ points, and they are joined by the connector.
Components beyond the second are excluded by construction.

The encoder is a fully connected network $320 \to 160 \to 40 \to 10 \to
\latdim$ with a mirrored decoder, dropout $0.1$, and Adam at a learning rate of
$10^{-4}$ with batch size $256$. The architecture is deliberately small: the
comparison of interest is between loss functions, and a high-capacity network
would let reconstruction succeed regardless of how the latent is organised,
which is precisely the effect under test.

The headline latent dimension is $\latdim = 2$, with $\latdim = 5$ reported
throughout as a robustness check. Two dimensions are chosen on the effective
dimensionality measured in \Cref{sec:res:dimension}, not on convenience; that
the choice costs nothing measurable is a result and is reported there rather
than assumed here.

\subsection{Regularisation}\label{sec:setup:reg}

The neighbour graph uses $k = 50$. The kernel is fitted from
$\texttt{min\_dist} = 0.01$ and $\texttt{spread} = 1.0$, giving the $(a,b)$ of
\Cref{eq:qij}; repulsion strength is $\nu = 1$ with $5$ negative samples per
edge.

Training follows the two-stage curriculum of \Cref{sec:method:loss}. A
reconstruction-only warm-up runs to convergence, up to $4{,}000$ epochs under
early stopping. The graph term is then introduced and $\lambda$ raised through
six phases,
\[
  \lambda:\; 10^{-7} \to 10^{-6} \to 10^{-5} \to 10^{-4} \to 10^{-3}
    \to 10^{-2},
\]
at $200$ epochs each for the first five and up to $3{,}000$ for the last. The
early phases are a fixed curriculum and run in full; the warm-up and the final
phase target convergence and stop early once the total loss has not improved by
$10^{-6}$ for $150$ epochs.

\Cref{sec:res:umap} reports what $\texttt{min\_dist}$ does across four settings,
and \Cref{sec:res:metric} what the graph metric does across five.

\subsection{Neighbour-graph metric}\label{sec:setup:metric}

The graph is built with Euclidean distance on \srvf{} coordinates. As
established in \Cref{sec:method:graph}, this is not a departure from the
geodesic: on unit-norm \srvf{}s the chord is a strictly increasing function of
the great-circle angle (\Cref{eq:chord-arc}), so the two induce the same
neighbour sets. Measured over the corpus, the rank correlation between the two
distances is $1.000000$.

The same holds for cosine distance, which is likewise strictly monotone in the
angle. None of the three is more faithful to the shape metric than the others,
and \Cref{sec:res:metric} shows the choice does not propagate to any measured
outcome.

\subsection{Interpolation protocol}\label{sec:setup:poc1}

The held-out-anchor protocol has one free parameter, the anchor step, and it is
consequential: at a step of $120$ roughly six anchors must generate $647$
frames, while at a step of $10$ the model still predicts $90\%$ of the sequence
from the remainder. Results are reported at a step of $10$ unless stated, with
the full density sweep in \Cref{tab:poc1-density}, because a single density
cannot distinguish a limitation of the model from a limitation of the sampling.

An anchor is additionally placed on each side of every topology change. Runs
between changes are often only a few frames long, so a regular grid drops none
into them and the piecewise scheme has nothing to fit; anchoring the boundaries
is also what an experimenter would do, since the transition is the event worth
measuring. A consequence is that held-out frames near a change are scored as
\emph{neighbours} of the discontinuity rather than as the discontinuity itself,
and the reported transition error measures how far the damage spreads either
side of a change rather than how badly the jump itself is predicted.

\subsection{Baselines}\label{sec:setup:baselines}

Two baselines, both trained on the same data with the same protocol.

\paragraph{POD.}
Proper orthogonal decomposition of the \srvf{} vectors: the linear
reduced-order model against which any nonlinear method should justify its
added complexity. It is deterministic and therefore quoted without a seed
spread.

\paragraph{Geometric-AE.}
An autoencoder with an isometry penalty on the decoder Jacobian,
\begin{equation}\label{eq:isoloss}
  \mathcal{L} = \mathcal{L}_{\mathrm{recon}}
    + \mu \bigl\lVert J^{\top} J - c I \bigr\rVert_F^2,
  \qquad c = \tfrac{1}{\latdim}\operatorname{tr}(J^{\top}J),
\end{equation}
with $\mu = 10^{-3}$, evaluated on $32$ latent points per batch. This drives the
decoder towards a local isometry up to a global scale, following
\citet{atzmon2020isometric} and the scaled-orthonormality argument of
\citet{kato2020rate}.

It is the natural competitor because it also aims at a geometrically
well-behaved latent, but by a different mechanism: a pointwise condition on the
decoder rather than a dataset-wide condition on neighbourhoods. Note that it is
distinct from the geometry-regularised autoencoder of
\citet{duque2022geometry}, which regularises the bottleneck with diffusion
potentials and is therefore aware of the data's neighbourhood structure;
\Cref{eq:isoloss} is not.

\subsection{Statistical protocol}\label{sec:setup:stats}

Every model is trained at three random seeds, and every reported value is the
mean over those seeds, quoted as mean $\pm$ standard deviation. The standard
deviation is the \emph{sample} standard deviation (denominator $n-1$), since
the seeds are a sample of the training distribution rather than the whole of
it; with $n = 3$ this is $1.22\times$ the population value, so the distinction
is not negligible and is applied uniformly. A difference between two
configurations is called significant only when it exceeds twice the
pooled standard deviation of the two.

This threshold is applied without exception, and it changes several
conclusions. Differences that appear decisive at a single seed --- including
some that would make a stronger paper --- do not survive it, and are reported
as null results in the same voice as the positives. Where a comparison fails
the test but shows a consistent ordering across seeds, it is described as a
trend and the pooled-sd ratio is given, so the reader can judge it directly.

Three seeds bounds what can be claimed. It establishes the absence of a
\emph{large} effect, not the absence of an effect, and where the seed spread is
comparable to the difference under test this is stated at the point of use.

\section{Results}\label{sec:results}

Whether a corpus dense in shape can supply what a recording sparse in time
omitted turns on one property: that position in the latent carries shape
meaning. The results are ordered to test that property and then to use it,
rather than to follow the pipeline.

\Cref{sec:res:ablation} establishes the effect itself, by removing the graph
term from an otherwise identical model. \Cref{sec:res:dimension} shows the
resulting latent is small enough to be drawn on a page, and
\Cref{sec:res:cleaning} what the data preparation contributes to that.
\Cref{sec:res:poc1,sec:res:poc2,sec:res:poc3} then ask what such a latent is
good for: predicting unobserved states in time, generating a full interface
from process parameters, and recovering morphological structure.
\Cref{sec:res:benchmark} places the method against a linear reduction and
against an autoencoder that pursues latent geometry by a different route.
\Cref{sec:res:variants,sec:res:metric,sec:res:umap} report what turned out not
to matter, at the same length as what did.

Every value is a mean over three random seeds with its sample standard
deviation, and a difference is called significant only when it exceeds twice
the pooled standard deviation of the two configurations compared
(\Cref{sec:setup:stats}).

\subsection{Ablation of the manifold-informed loss}\label{sec:res:ablation}

Every configuration described so far carries the graph term of
\Cref{eq:loss}. The first question is therefore whether that term does
anything at all. The ablation removes it and changes nothing else: identical
architecture, data, schedule and seed, with $\lambda$ held at zero throughout
(\Cref{tab:ablation}).

\begin{table}[t]
  \centering
  \caption{Ablation of the manifold-informed loss. Identical architecture,
  data, schedule and seed; the only difference is the regularisation term.
  Three seeds, mean $\pm$ sd.}
  \label{tab:ablation}
  \begin{tabular}{lccc}
    \toprule
    & Plain AE (V0) & \shrom{} (V2) & pooled sd \\
    \midrule
    ARI, $K=4$, $\latdim=5$   & $0.063 \pm 0.059$ & $\mathbf{0.781 \pm 0.049}$ & $\mathbf{13.2}$ \\
    ARI, $K=4$, $\latdim=2$   & $0.085 \pm 0.054$ & $\mathbf{0.829 \pm 0.059}$ & $\mathbf{13.2}$ \\
    ARI, $K=8$, $\latdim=5$   & $0.092 \pm 0.044$ & $\mathbf{0.569 \pm 0.023}$ & $13.6$ \\
    PoC-2 $R^2$, $\latdim=5$  & $0.748 \pm 0.142$ & $0.868 \pm 0.010$ & $0.6$ \\
    PoC-1 @120, $\latdim=5$   & $\mathbf{0.123 \pm 0.005}$ & $0.136 \pm 0.010$ & $1.1$ \\
    \bottomrule
  \end{tabular}
\end{table}

An adjusted Rand index of zero is chance-level agreement and one is an exact
match (\Cref{tab:scales}). At
$\mathrm{ARI} = 0.063 \pm 0.059$ the plain autoencoder is not distinguishable
from chance: clustering its latent recovers essentially nothing of the
partition that clustering the shapes themselves produces. With the graph term
the same architecture reaches $0.781 \pm 0.049$, a separation of $13.2$ pooled
standard deviations, and $0.829 \pm 0.059$ at $\latdim = 2$. These are the
largest effects in the study.

\begin{figure}[t]
  \centering
  \includegraphics[width=\textwidth]{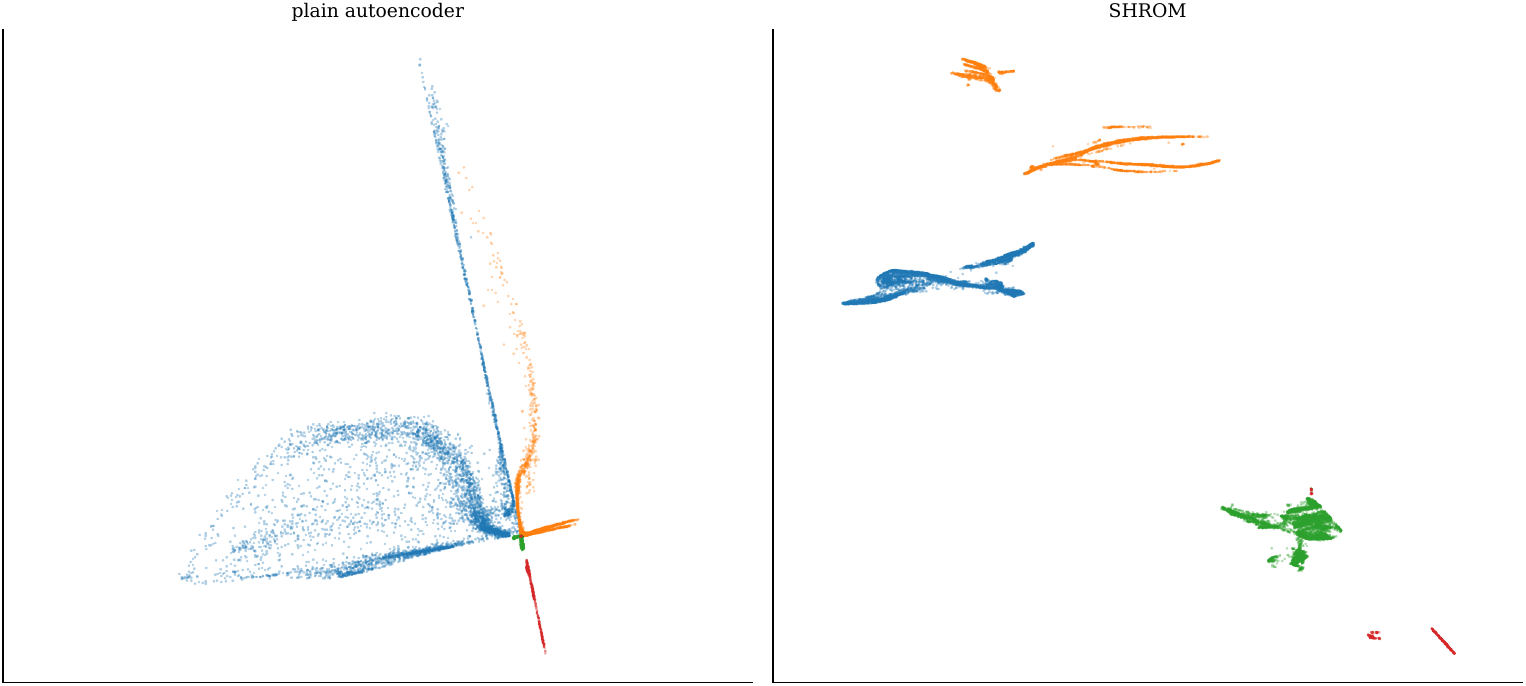}
  \caption{The ablation seen in the latent itself. Both panels show the same
  shapes, coloured by the same partition of shape space (geodesic $k$-means,
  $K=4$), so neither model is scored on its own terms. The plain autoencoder
  (left) does separate the classes, but joins them: its clusters meet along
  filaments at a central hub, so a path between two regions stays over occupied
  latent. \shrom{} (right) resolves them into four islands with empty space
  between. The right panel is what recovers the shape-space partition
  (\Cref{tab:ablation}); the left is what makes interpolation across a class
  boundary easier (\Cref{sec:res:poc1}). The two proofs of concept want
  opposite things from the same property.}
  \label{fig:latent-compare}
\end{figure}

\Cref{fig:latent-compare} shows the same ablation in the latent rather than in
decoded shapes, and corrects a reading the table invites: the plain
autoencoder's latent is not disordered, it is \emph{connected}. Its clusters
are separated but joined, where \shrom{}'s are separated and disjoint. That
distinction is what the next subsections turn on.

\paragraph{Improvement is confined to the expected proof of concept.}
Regression and interpolation are unchanged within noise, and at $\latdim = 5$
the plain autoencoder is nominally \emph{better} at interpolation
($0.123 \pm 0.005$ against $0.136 \pm 0.010$, $1.1$ pooled sd). This is not a
weakness of the ablation but a consequence of what each task measures.

Interpolation and regression \emph{read out} the latent through a fitted map.
A readout can compose with any invertible reparametrisation of the latent
coordinates, so it is largely indifferent to how the space is organised: a
badly arranged latent that is nonetheless injective can be read out about as
well as a well-arranged one. Clustering, grid decoding and structure overlap
ask a different question --- whether \emph{position} in the latent carries
meaning --- and that is the property a generative model depends on. It is also
where the graph term makes an $11$ to $16$ standard-deviation difference.

\begin{figure}[t]
  \centering
  \includegraphics[width=0.72\textwidth]{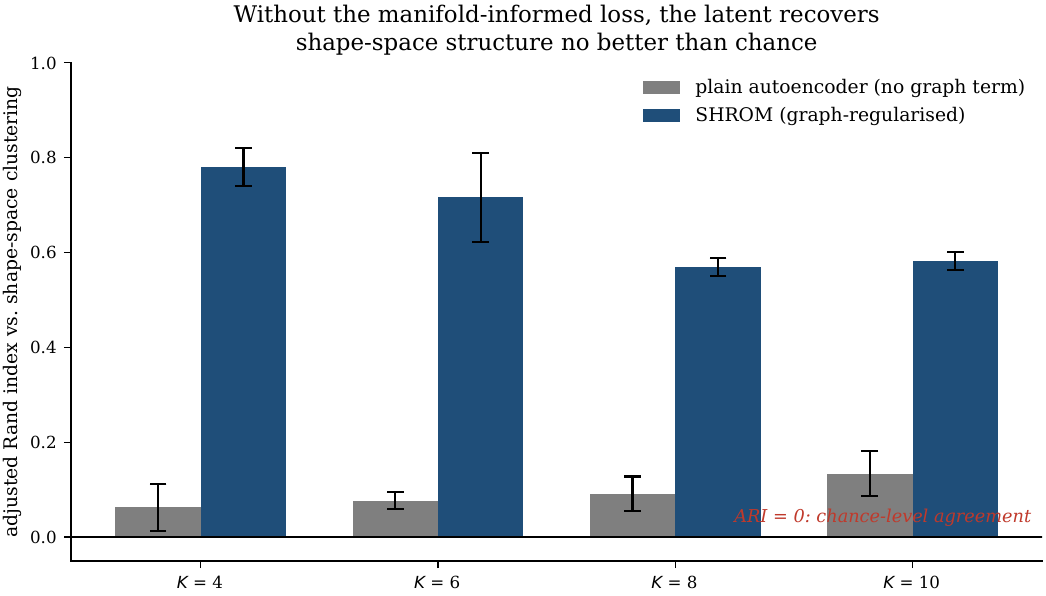}
  \caption{\Cref{tab:ablation} as a chart. An ARI of zero is chance-level
  agreement; the plain autoencoder does not clear it at any $K$.}
  \label{fig:ablation-bars}
\end{figure}

\begin{figure}[t]
  \centering
  \includegraphics[width=\textwidth]{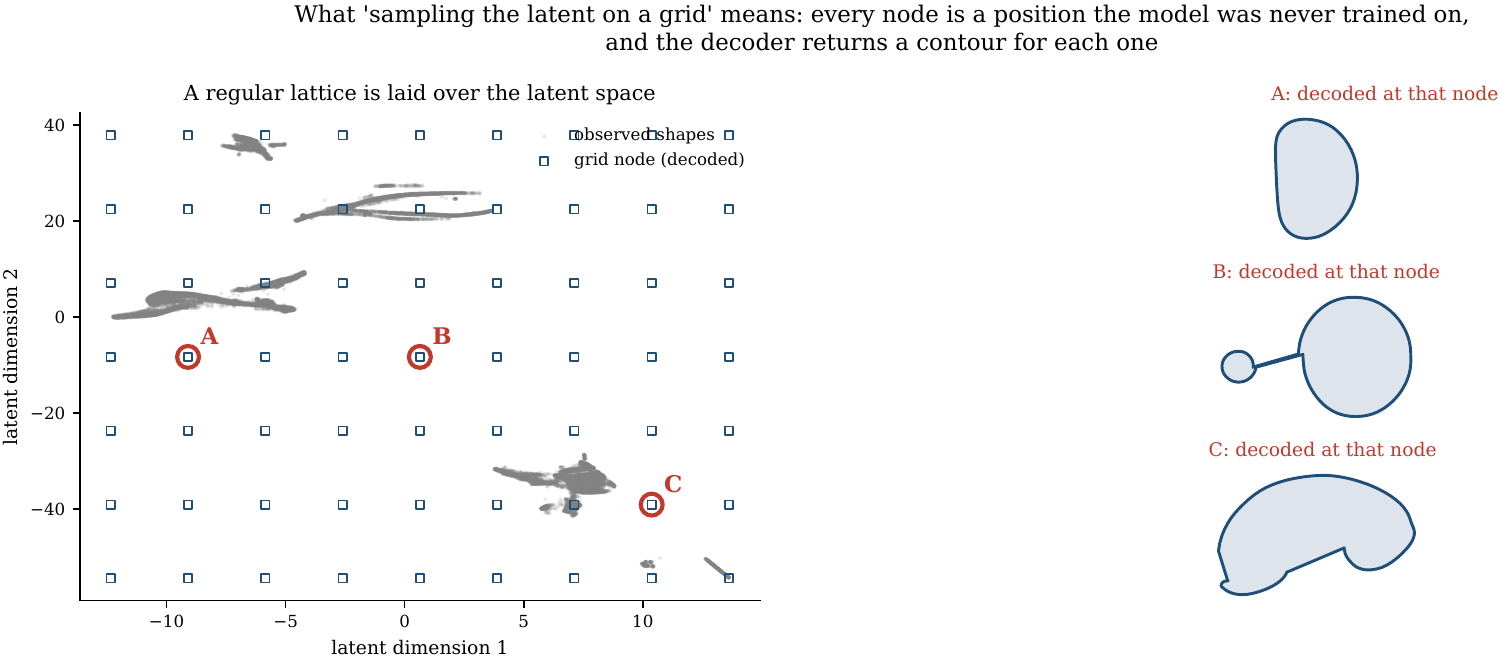}
  \caption{What grid sampling means. A regular lattice is laid over the latent
  space (left); each node is a coordinate the model never saw in training, and
  the decoder returns a contour for it (right, three marked nodes). The grid
  figures that follow show every node of such a lattice.}
  \label{fig:grid-explainer}
\end{figure}

\begin{figure}[t]
  \centering
  \includegraphics[width=\textwidth]{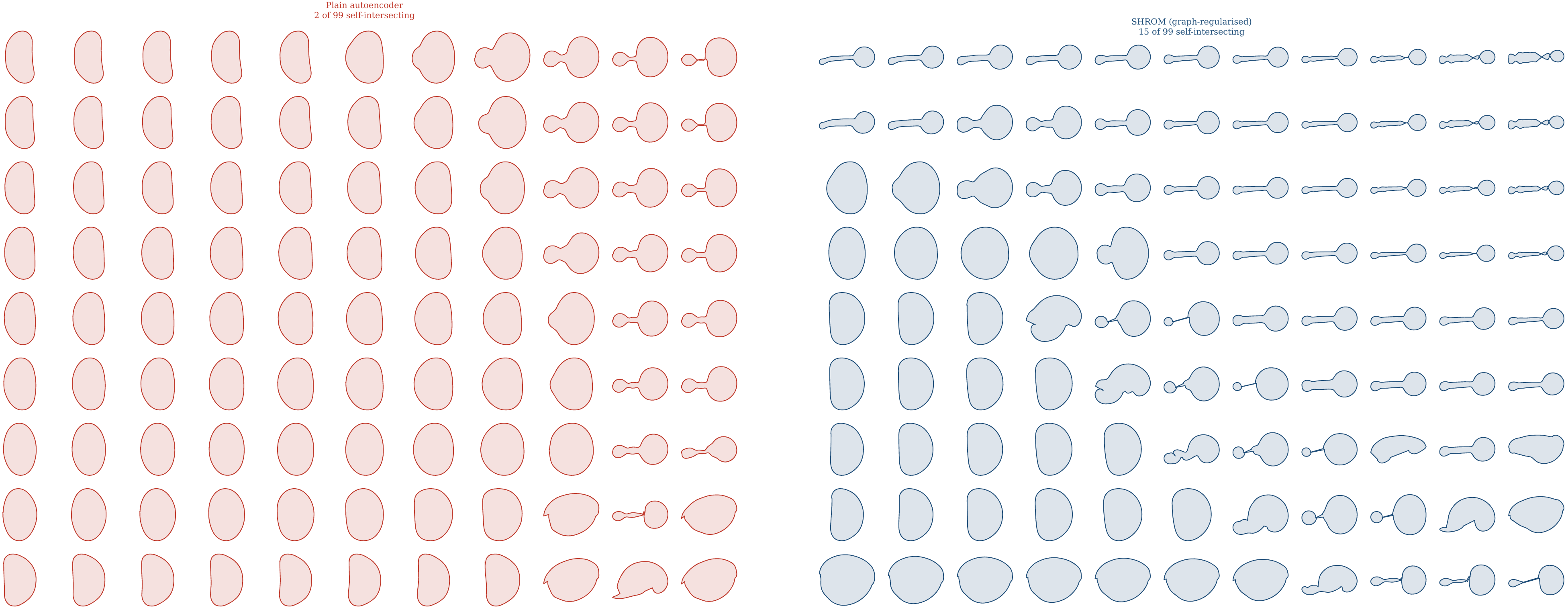}
  \caption{The ablation made visible. Decoded shapes on a regular grid over the
  two-dimensional latent; every node is a position never seen in training.
  Left: plain autoencoder. Right: \shrom. The plain model produces fewer
  self-intersecting outlines ($2\%$ against $16\%$), but only because $80\%$
  of its grid is the same near-circular oval (solidity median $0.999$, sd
  $0.076$); \shrom{} spans ligaments, necking and satellites, with
  $2.2\times$ the solidity spread. The $16\%$ is the cost of that coverage and
  occurs where the latent crosses between topology classes.}
  \label{fig:grid-ablation}
\end{figure}

\subsection{Latent dimensionality}\label{sec:res:dimension}
Two statements, established independently.

First, the latent \emph{uses} no more than two dimensions. The participation
ratio of the decoder-Jacobian singular values is $n_{\mathrm{eff}} = 1.20 \pm
0.02$ at $\latdim = 5$, indicating that one direction dominates locally, and
$99.2\%$ of the latent variance lies in two principal components. The two
statistics measure different things --- the first is local and derived from the
decoder, the second global and derived from the latent codes --- and together
they bound the used dimension between one and two. A five-dimensional latent is
available to the model; it declines most of it.

Second, two dimensions are \emph{enough}. \Cref{tab:dimension} compares the
headline model at $\latdim = 2$ and $\latdim = 5$ across every task. No
comparison reaches significance, and $\latdim = 2$ is nominally ahead on three
of the four.

Neither statement is a claim about the intrinsic dimension of the shape data,
which is a property of the point cloud rather than of any fitted model, and
which we measure separately. Three estimators applied to the \srvf{} corpus
disagree in an informative way. The correlation dimension is $1.9$ and is stable
at every sample size tested. The TwoNN estimator \citep{facco2017twonn} rises
from $9.2$ to $11.8$ as the sample grows from $10{,}000$ to $40{,}000$ points,
and the maximum-likelihood estimator of \citet{levina2004maximum} falls from
$9.7$ to $7.1$ as the neighbourhood size $k$ grows from $5$ to $25$. Drift of
this kind indicates structure at more than one scale: the corpus is close to a
low-dimensional surface when viewed coarsely, and acquires further degrees of
freedom when resolved finely. A global principal-component analysis agrees, with
two components carrying $85\%$ of the variance and eight required for $95\%$.

The data manifold is therefore not two-dimensional, and we do not claim that it
is. What the measurements support is the narrower statement made above: the
trained decoder uses about two latent directions, and two suffice for every
downstream task reported here. A representation can compress well below the
intrinsic dimension of its data whenever the discarded directions do not affect
the quantities being predicted, which is what \Cref{tab:dimension} tests
directly.

That reasoning invites a direct test: train at the measured intrinsic
dimension. \Cref{tab:highdim} reports \shrom{} at $\latdim = 8$ and
$\latdim = 10$, three seeds each, otherwise identical to the headline model.
Interpolation error is $0.0902 \pm 0.0064$ and $0.0954 \pm 0.0157$ against
$0.0972 \pm 0.0085$ at $\latdim = 2$; both are nominally ahead, and the
largest separation is $0.9$ pooled standard deviations. The waveform regression
and the clustering agreement behave the same way, with every comparison against
$\latdim = 2$ inside twice the pooled standard deviation. The latent dimension
can therefore be raised to the estimated intrinsic dimension of the data
without a measurable effect on any of the three tasks. Decoding a sweep along each latent coordinate shows why:
at $\latdim = 8$ roughly three coordinates produce visible changes in
morphology and the remainder leave the decoded shape nearly fixed, which is
$n_{\mathrm{eff}}$ in visual form.

\begin{figure}[t]
  \centering
  \includegraphics[width=0.86\textwidth]{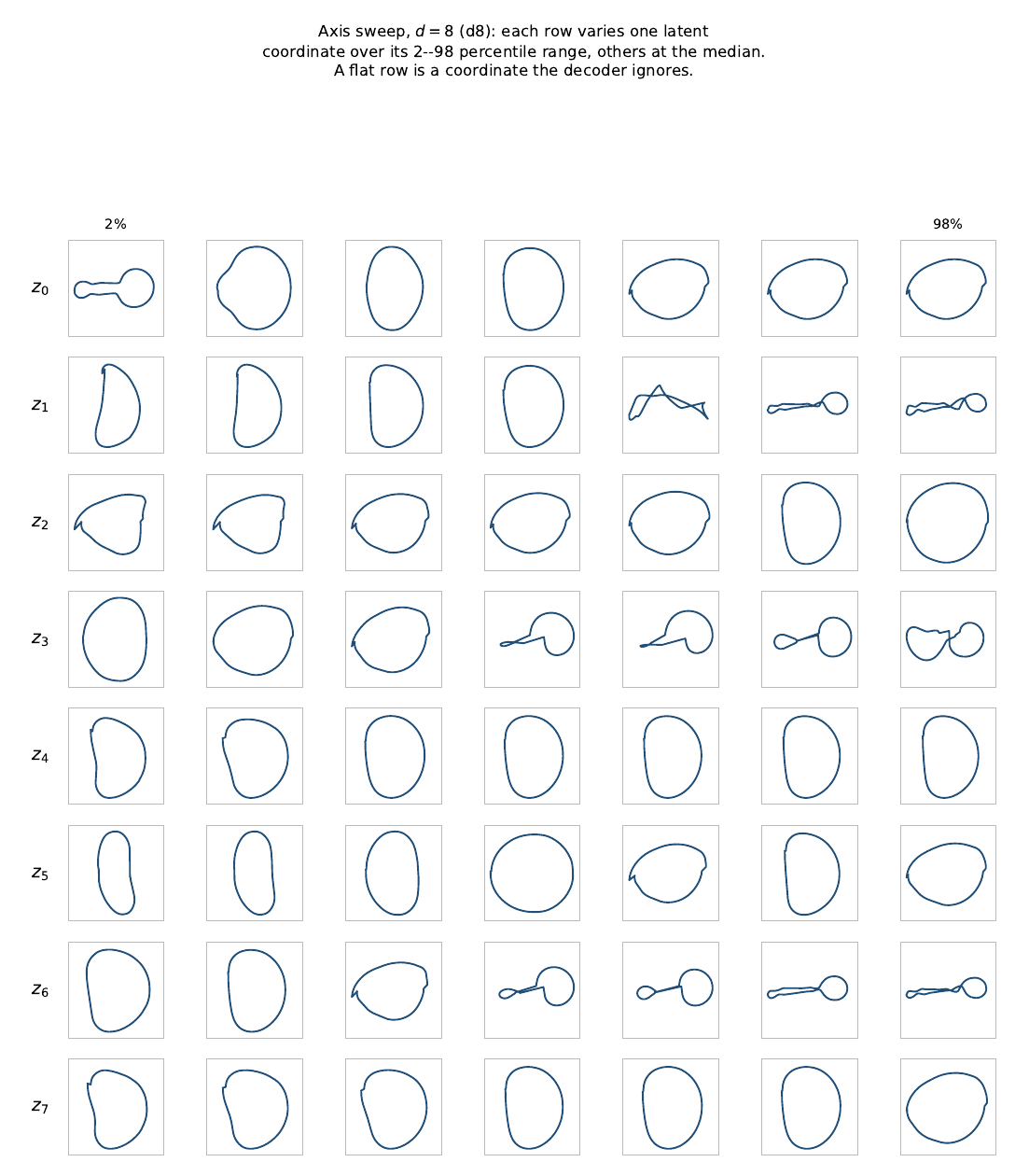}
  \caption{Decoded sweep along each latent coordinate at $\latdim = 8$. Each
  row varies one coordinate across its $2$--$98$ percentile range with the
  others held at the data median; no projection or dimensionality reduction is
  applied. Three coordinates drive visible changes in morphology --- ligaments,
  necking, satellite separation --- and the remainder leave the decoded shape
  nearly fixed. The model declines most of the capacity it is given, which is
  what $n_{\mathrm{eff}}$ reports as a number.}
  \label{fig:axissweep-d8}
\end{figure}

\begin{table}[t]
  \centering
  \caption{Latent dimension against every downstream task, mean $\pm$ sample
  standard deviation over three seeds. PoC-1 is interpolation error at anchor
  step $10$ (lower is better); PoC-2 is waveform-to-contour $R^2$ and PoC-3 the
  adjusted Rand index at $K=4$ (higher is better). No comparison against
  $\latdim = 2$ reaches twice the pooled standard deviation.}
  \label{tab:highdim}
  \begin{tabular}{lccc}
    \toprule
    Latent dimension & PoC-1 geodesic & PoC-2 $R^2$ & PoC-3 ARI \\
    \midrule
    $\latdim = 2$  & $0.0972 \pm 0.0085$ & $0.878 \pm 0.026$ & $0.829 \pm 0.059$ \\
    $\latdim = 5$  & $0.1005 \pm 0.0138$ & $0.868 \pm 0.010$ & $0.781 \pm 0.049$ \\
    $\latdim = 8$  & $\mathbf{0.0902 \pm 0.0064}$ & $0.857 \pm 0.016$ & $0.873 \pm 0.022$ \\
    $\latdim = 10$ & $0.0954 \pm 0.0157$ & $\mathbf{0.883 \pm 0.024}$ & $\mathbf{0.882 \pm 0.027}$ \\
    \bottomrule
  \end{tabular}
\end{table}

\begin{table}[t]
  \centering
  \caption{Dimension study, headline model, three seeds.}
  \label{tab:dimension}
  \begin{tabular}{lccc}
    \toprule
    Metric & $\latdim=2$ & $\latdim=5$ & verdict \\
    \midrule
    PoC-2 $R^2$      & $\mathbf{0.878 \pm 0.026}$ & $0.868 \pm 0.010$ & $0.4$\,sd \\
    ARI $K=4$        & $\mathbf{0.829 \pm 0.059}$ & $0.781 \pm 0.049$ & $0.9$\,sd \\
    ARI $K=8$        & $0.502 \pm 0.046$ & $\mathbf{0.569 \pm 0.023}$ & $1.8$\,sd \\
    PoC-1 @120       & $\mathbf{0.120 \pm 0.001}$ & $0.136 \pm 0.010$ & $1.6$\,sd \\
    reconstruction   & $2.07\times10^{-4}$ & $\mathbf{1.22\times10^{-4}}$ & the one cost \\
    \bottomrule
  \end{tabular}
\end{table}

The order of these two statements matters. The latent dimension was not
selected by tuning on the metrics in \Cref{tab:dimension}: $n_{\mathrm{eff}}$
was measured first, indicated that two dimensions would suffice, and the table
then confirmed it. Selecting $\latdim$ by scanning downstream scores would make
that table a fit rather than a test.

There is one real cost. Reconstruction is about $1.7\times$ better at
$\latdim = 5$ ($1.22 \times 10^{-4}$ against $2.07 \times 10^{-4}$; these are
mean squared errors on unit-norm vectors, so only the ratio carries meaning),
and this is the only metric on which the larger latent wins. No proof of concept
required that additional fidelity, which is consistent with the reconstruction
argument of \Cref{sec:res:benchmark}: reconstruction error and latent
organisation are close to independent.

\subsection{Effect of data cleaning on the latent}\label{sec:res:cleaning}
\begin{figure}[t]
  \centering
  \begin{subfigure}{0.48\textwidth}
    \includegraphics[width=\textwidth]{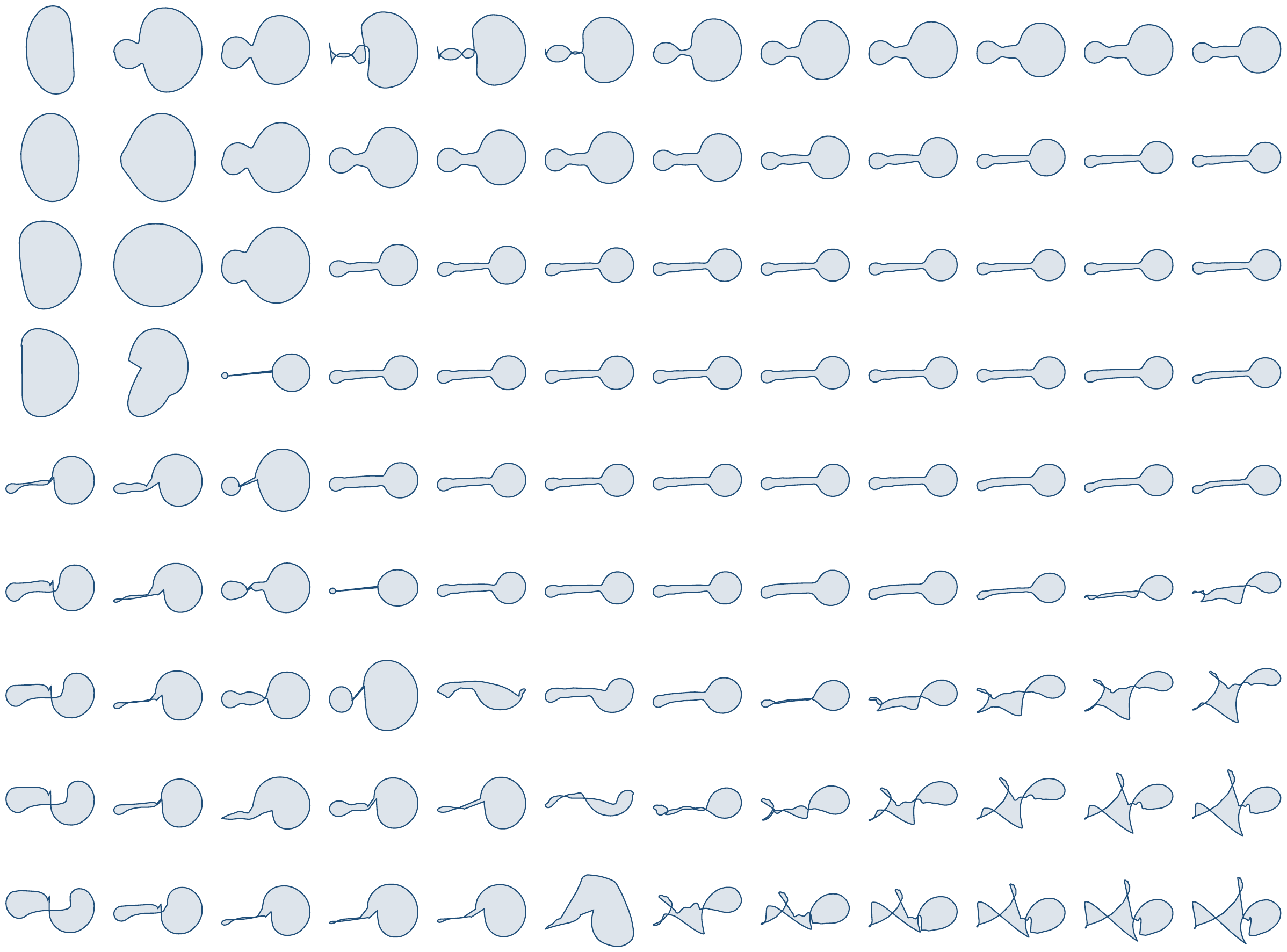}
    \caption{One-stage cleaning}
  \end{subfigure}\hfill
  \begin{subfigure}{0.48\textwidth}
    \includegraphics[width=\textwidth]{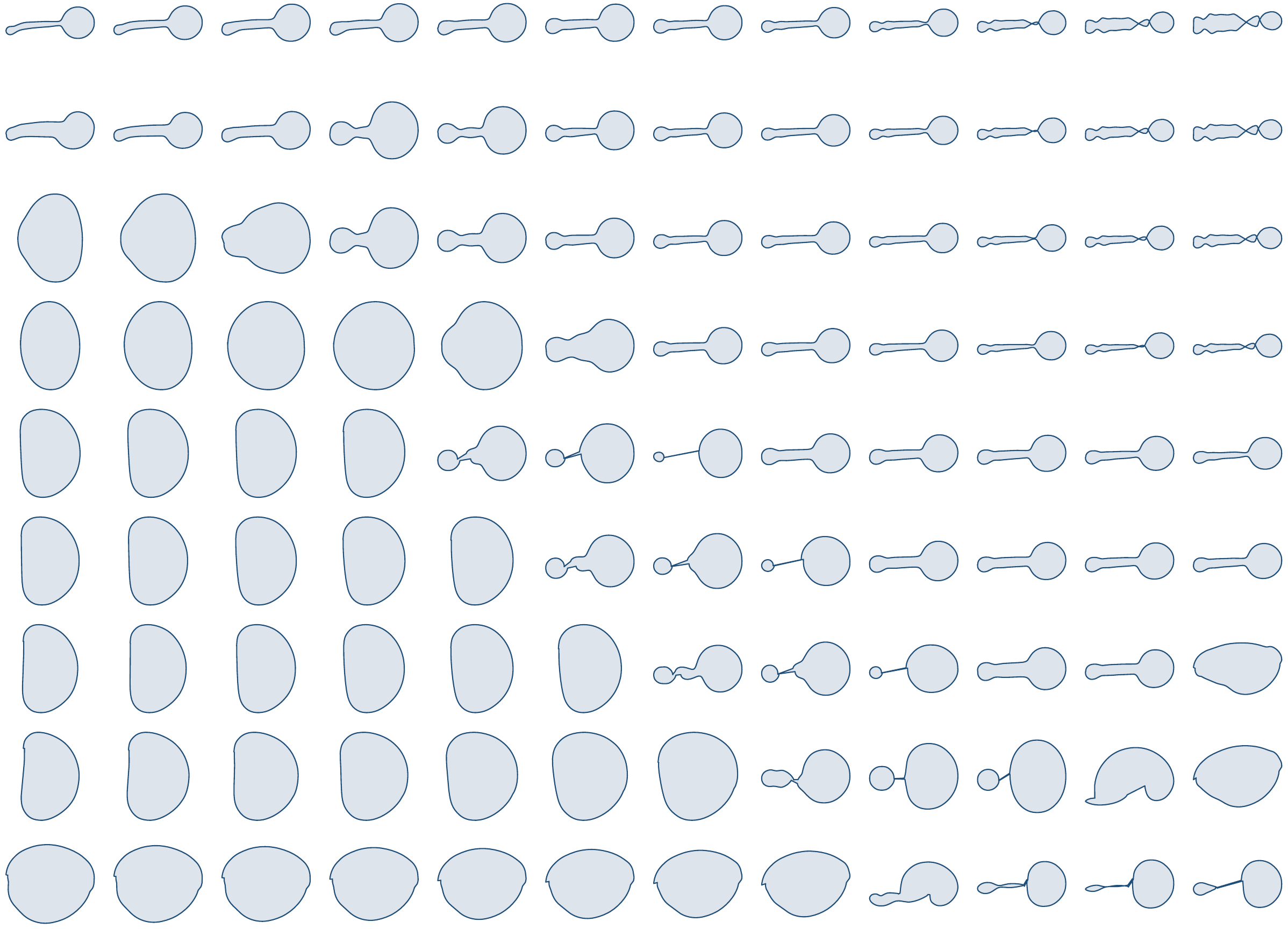}
    \caption{Two-stage cleaning}
  \end{subfigure}
  \caption{Effect of stage-2 filtering on latent decodability. Decoded shapes
  on a regular grid over the latent, before and after the per-cluster outlier
  removal of \Cref{sec:setup:cleaning}.}
  \label{fig:grids-cleaning}
\end{figure}

Stage-2 filtering removes $1.00\%$ of shapes
(\Cref{sec:setup:cleaning}) and changes the latent qualitatively. Before it,
one quadrant of the decoded grid is unusable; after it, all $216$ grid nodes
decode to plausible droplets (\Cref{fig:grids-cleaning}). Latent anisotropy
rises from $3.2$ to $33.8$, concentrating the space onto a single ordered
morphology axis rather than spreading it across directions that carry little
shape variation.

The scalar metrics do not improve. Reconstruction, regression and
clustering are all unchanged within seed noise. The benefit is
representational: it appears in what the latent decodes to, not in what the
summary statistics report. For a model whose purpose is to be sampled at unobserved coordinates, that is
the relevant axis. No scalar reported here captures it, so
\Cref{fig:grids-cleaning} carries this result in place of a table.

\begin{figure}[t]
  \centering
  \includegraphics[width=0.7\textwidth]{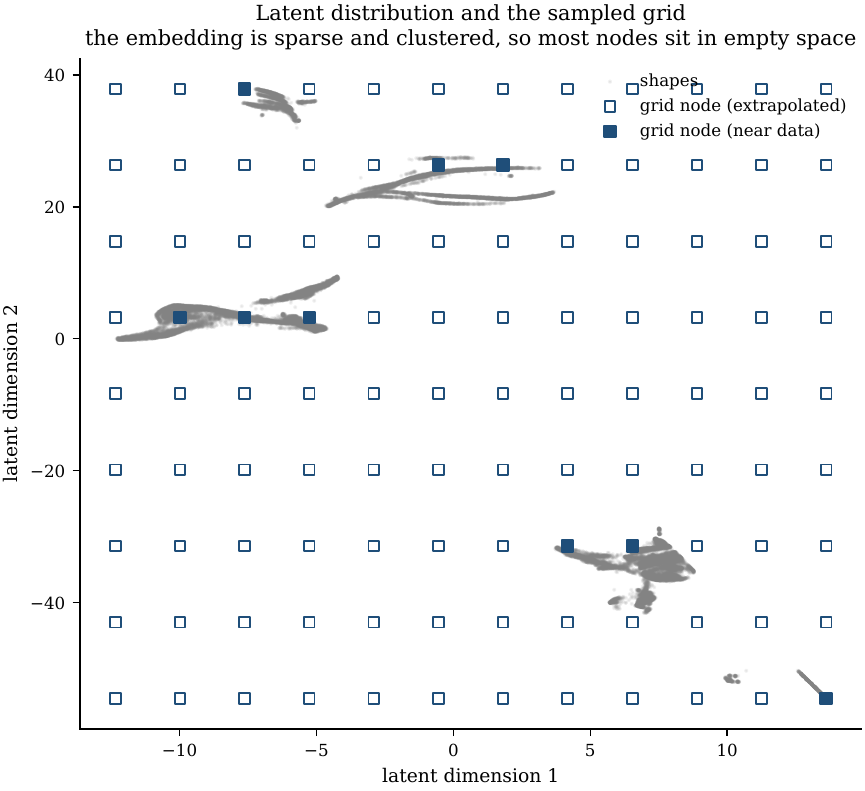}
  \caption{The latent point cloud with the sampled grid nodes marked. Nodes
  falling inside the cloud are interpolation; those outside it are
  extrapolation, and are decoded on the same terms.}
  \label{fig:latent-cloud}
\end{figure}

\subsection{PoC-1: temporal interpolation}\label{sec:res:poc1}
Both protocols of \Cref{sec:method:protocol} are reported, because they answer
different questions and differ substantially in difficulty. The held-out-anchor
protocol fits a path through a sparse subset of an unseen acquisition and
evaluates at the withheld frames; the leave-one-out protocol withholds a single
frame from an otherwise complete series. The latter is the standard
non-intrusive ROM protocol and is about $1.5\times$ easier here. Reporting only
one would either overstate or understate what the model does.

\paragraph{Error reporting by regime.}
A droplet spends most of a sequence in a relaxed state and a small fraction
actively deforming. Interpolation error differs sharply between the two:
$0.325$~rad in the deforming phase against $0.064$~rad when relaxed, a
$5.1\times$ gap that is consistent across all four trajectories. For scale
(\Cref{tab:scales}), two shapes drawn at random from the corpus differ by
$1.06$~rad, so even the deforming-phase error is under a third of the
separation between two arbitrary droplets, and the relaxed-phase error is
about a sixteenth of it. Because the deforming phase is
only about $10\%$ of frames, any pooled mean is dominated by the easy regime
and understates the difficulty of precisely the physics of interest. All
interpolation results are therefore stratified by regime.

\begin{table}[t]
  \centering
  \caption{Interpolation schemes at anchor step 120, three seeds. Geodesic
  distance to the withheld frame.}
  \label{tab:poc1}
  \begin{tabular}{lccc}
    \toprule
    Scheme & \shrom{} (V2) & V5 & V6 \\
    \midrule
    piecewise      & $\mathbf{0.136 \pm 0.010}$ & $0.143 \pm 0.010$ & $\mathbf{0.114 \pm 0.003}$ \\
    SRVF reference & $0.141 \pm 0.015$ & $0.149 \pm 0.016$ & $0.121 \pm 0.004$ \\
    monotone       & $0.145 \pm 0.003$ & $0.161 \pm 0.012$ & $0.136 \pm 0.007$ \\
    latent geodesic& $0.222 \pm 0.124$ & $0.152 \pm 0.011$ & $0.156 \pm 0.046$ \\
    naive spline   & $0.943 \pm 0.033$ & $0.819 \pm 0.077$ & $0.927 \pm 0.214$ \\
    \bottomrule
  \end{tabular}
\end{table}

Two claims follow from \Cref{tab:poc1}.

First, the naive cubic spline fails decisively --- $5.7$ to $8.1\times$ worse
than any topology-aware scheme. It fits a smooth curve through latent codes
that straddle a discontinuity, rings across it, and decodes to
self-intersecting curves. The failure is systematic rather than marginal, and
it is the clearest evidence that latent paths cannot be fitted without regard
to topology.

Second, the four topology-aware schemes are statistically
\emph{indistinguishable}. The piecewise scheme is lowest in all nine
model-scheme comparisons, which is a consistent ordering worth reporting, but
no individual gap clears twice the pooled standard deviation. We therefore
report the consistency and not the magnitude, and do not claim a best scheme.

\paragraph{Variation across sequences.}
Pooling the four filmed sequences hides a factor of two between them, and the
spread is a property of the data rather than of the model
(\Cref{tab:poc1-traj}). The hardest sequence carries seven transition frames in
the shortest usable window; the easiest contains no topology change at all.
Every variant we trained ranks them in the same order, so the difficulty is
intrinsic to the sequence.

\begin{table}[t]
  \centering
  \caption{Interpolation error by sequence at anchor step $10$, piecewise
  scheme. The pooled mean is $0.092$; the hardest sequence is more than twice
  the easiest, and is also the one with the most break-up events in the fewest
  frames.}
  \label{tab:poc1-traj}
  \begin{tabular}{lccc}
    \toprule
    Sequence & Held-out frames & Transition frames & Geodesic error \\
    \midrule
    A & $800$ & $0$ & $0.067$ \\
    B & $573$ & $4$ & $0.070$ \\
    C & $386$ & $7$ & $0.151$ \\
    D & $487$ & $4$ & $0.079$ \\
    \bottomrule
  \end{tabular}
\end{table}

We report all four throughout. Excluding the hardest would improve every
aggregate in this paper, and choosing to do so after seeing which sequence is
hardest would not be a defensible protocol.

\paragraph{Composition of the error.}
The anchor step is a free parameter of the protocol, not of the model, and the
error depends on it strongly: \Cref{tab:poc1-density} sweeps it from one anchor
in $120$ frames down to one in two.

\begin{table}[t]
  \centering
  \caption{Interpolation error against anchor density, piecewise scheme,
  averaged over the four sequences, single seed per row. The error falls
  steeply while anchors are scarce and then saturates: half the trajectory buys
  little over a tenth of it. The anchor-$10$ row is reproduced over three seeds
  in \Cref{tab:poc1-headtohead}, where the ordering holds in every seed.}
  \label{tab:poc1-density}
  \begin{tabular}{lccc}
    \toprule
    Anchor step & Frames withheld & \shrom{} & plain AE \\
    \midrule
    $120$ & $99\%$ & $0.128$ & $0.125$ \\
    $40$  & $97\%$ & $0.102$ & $0.088$ \\
    $10$  & $90\%$ & $0.092$ & $0.076$ \\
    $2$   & $50\%$ & $0.089$ & $0.072$ \\
    \bottomrule
  \end{tabular}
\end{table}

The curve saturates rather than falling to zero, and it saturates at a
value the interpolation cannot reach past. Three independent observations
identify what sets that floor. At the densest setting the four topology-aware
schemes agree to three decimals, and the \srvf{} reference --- which
interpolates between \emph{decoded} anchors and never touches the latent path
--- matches the best of them ($0.0885$ against $0.0891$). Leave-one-out, which
withholds a single frame from an otherwise complete sequence, gives $0.100$:
no better than withholding half of them.

\paragraph{Isolation by adjacent-frame interpolation.}
Joining two \emph{consecutive} frames by a straight line in the latent removes
the fitting problem entirely: with two points a straight line is the exact
interpolant, with no spline, no anchor spacing and no freedom to tune. Whatever
error remains is not the fit.

Away from a topology change, the shape decoded halfway along that line is
\emph{closer} to the true droplet than the endpoints' own reconstructions are
--- a ratio of $0.91$ to $0.98$ across every configuration and sequence tested.
Interpolating between adjacent frames costs nothing measurable. Across
a breakup the midpoint error rises to $1.5$ to $1.8$ times the reconstruction
error, but the true separation between those two frames is $1.08$~rad and even
the worst midpoint is under half of it: the line does not leave the data
manifold, it merely cannot represent a discontinuity, which no continuous path
can.

PoC-1 therefore measures reconstruction fidelity, not interpolation
quality, and the plain autoencoder leads it for the same reason it leads the
benchmark of \Cref{sec:res:benchmark}: reconstruction is what \shrom{} trades
for latent organisation. This proof of concept supports a narrower claim: that the latent admits
interpolation at arbitrary temporal resolution and returns plausible droplets at
every intermediate time, which the naive spline fails to do. It does not support
the claim that \shrom{} reconstructs withheld frames more accurately than a
model optimised for reconstruction alone.

\begin{figure}[t]
  \centering
  \includegraphics[width=\textwidth]{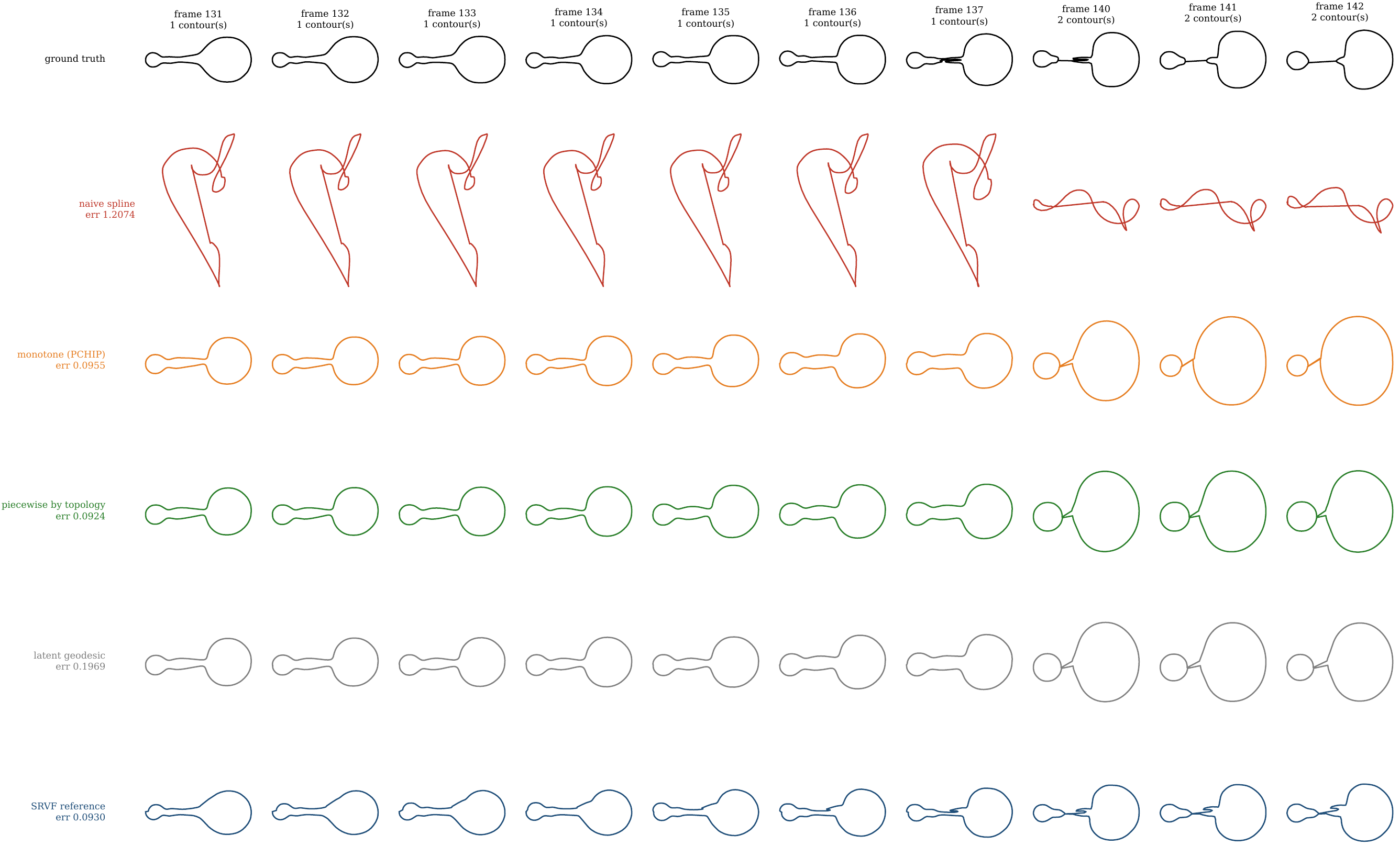}
  \caption{Ten consecutive held-out frames spanning a topology change (a single
  contour through frame~137, two contours from frame~140), with the ground
  truth above every interpolation scheme. At anchor step $120$ the interpolant
  sees $12$ of $653$ frames. The naive cubic spline fails outright, ringing
  across the discontinuity and decoding to self-intersecting curves, while the
  four topology-aware schemes remain plausible throughout and are statistically
  indistinguishable from one another.}
  \label{fig:poc1-evolution}
\end{figure}

\begin{figure}[t]
  \centering
  \includegraphics[width=\textwidth]{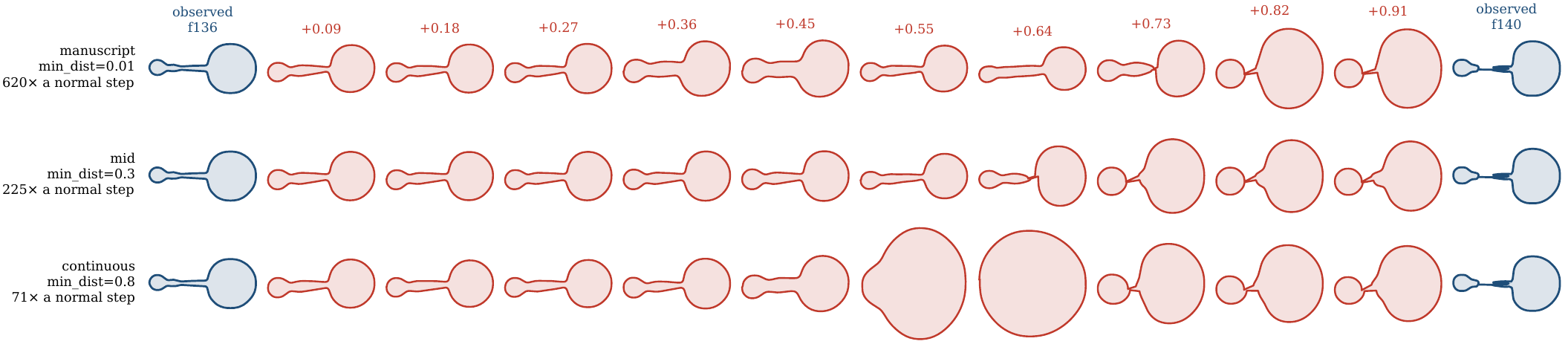}
  \caption{The pinch-off interval (frames~136 to~140, across which the
  component count changes) predicted at ten intermediate times, under three
  regularisation strengths. Blue endpoints are captured frames; the red shapes
  are times the camera never observed, and have no ground truth by
  construction. Row labels give the latent jump across the interval as a
  multiple of a typical consecutive step: $620\times$, $225\times$,
  $71\times$. The regulariser compresses the discontinuity by an order of
  magnitude without removing it. The strongest setting gives both the smallest
  jump and the \emph{worst} path through it --- it is the only row passing
  through shapes with no tail.}
  \label{fig:pinchoff}
\end{figure}

\paragraph{Failure regime.}
Within a topology class the model interpolates plausibly at arbitrary sub-frame
resolution: \Cref{fig:pinchoff} predicts ten intermediate states between two
captured frames, and the sequence remains a credible droplet throughout. Across
a change in component count it does not.

The latent jump across the pinch-off interval is $620\times$, $225\times$ and
$71\times$ a typical consecutive step under the three regularisation strengths.
Raising the regularisation compresses the discontinuity by nearly an order of
magnitude and does not remove it. The strongest setting is also instructive in
the wrong direction: it gives the smallest jump and the least plausible path
through it, being the only one that passes through shapes with no tail at all.
What this implies about the representation is taken up in
\Cref{sec:disc:limitation}.

\subsection{PoC-2: waveform to full contour}\label{sec:res:poc2}
A small network maps the eight waveform parameters to the latent code, from
which the decoder returns a full closed contour. Over three seeds at
$\latdim = 2$ this reaches $R^2 = 0.878 \pm 0.026$, with per-dimension values
between $0.82$ and $0.93$: every latent coordinate is predictable, rather than
one dominant coordinate carrying the score while the others are guessed. No
variant is separable from any other on this task.

The novelty is in the output, not the accuracy. Prior data-driven inkjet
surrogates predict scalar quantities --- volume, velocity, satellite count ---
or classify a regime. Droplet geometry has been predicted from physical inputs
elsewhere: \citet{wang2025pinchoff} relate the shape of a dripping drop at
pinch-off to fluid properties and flow rate, in both directions, using contours
and an autoencoder with latent clustering. Their inputs are material and flow
properties in a dripping rig rather than an actuation waveform in a
drop-on-demand device, and we are not aware of work predicting full closed
interface geometry from \emph{actuation parameters}, which is what the
composition with the decoder provides here.

\begin{figure}[t]
  \centering
  \includegraphics[width=0.72\textwidth]{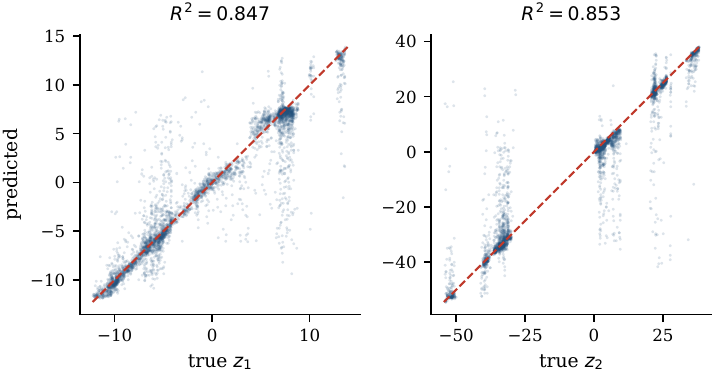}
  \caption{Predicted against true latent coordinate on the held-out test set
  ($N = 45\,230$), one panel per latent dimension; the dashed line is the
  identity. The cloud stays on the diagonal across the full operating range of
  each coordinate, so the mapping is well-conditioned rather than accurate only
  near the mean. This figure comes from a shorter training run than the
  headline number ($800$ epochs, single seed, $R^2 = 0.847$ and $0.853$); the
  reported $0.878 \pm 0.026$ is from the full schedule over three seeds.}
  \label{fig:poc2-scatter}
\end{figure}

\begin{figure}[t]
  \centering
  \includegraphics[width=\textwidth]{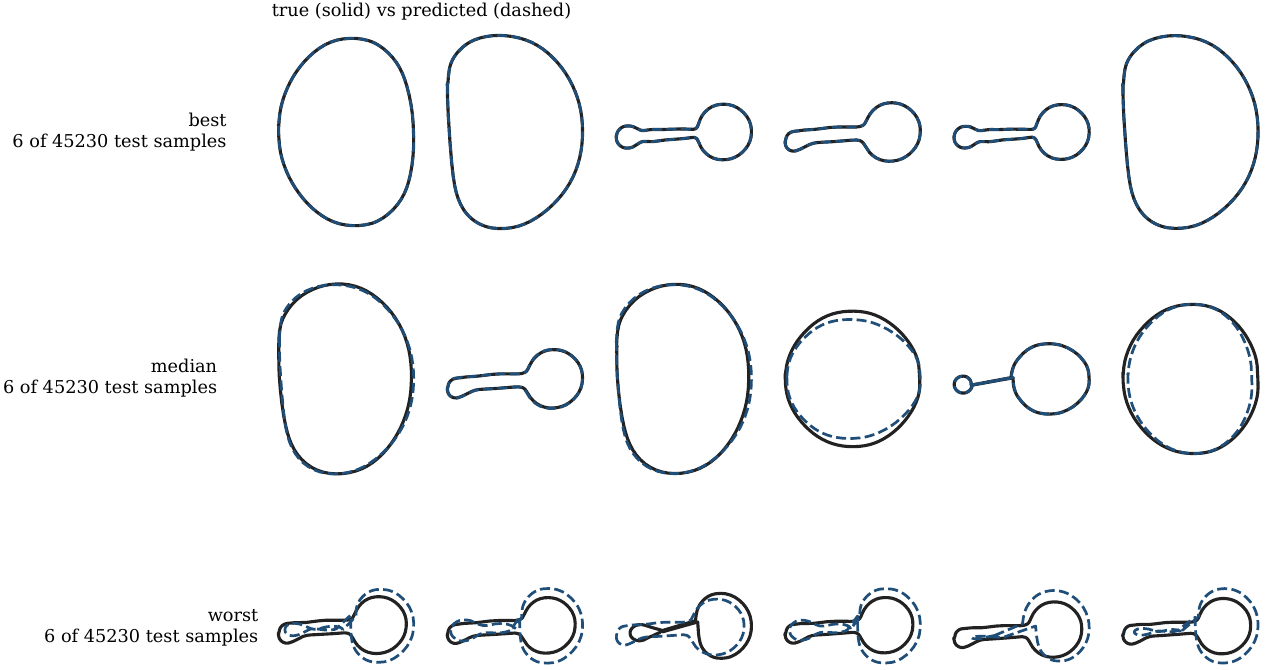}
  \caption{True (solid) against predicted (dashed) contours, decoded from the
  true and predicted latent codes respectively. Samples are \emph{stratified}
  by test-set error --- the six best, six median and six worst of $45\,230$ ---
  rather than drawn at random, so the failure modes are shown rather than
  averaged away. Even in the worst stratum the predicted shape retains the
  correct topology and tail: the error is in body size, not in morphology.}
  \label{fig:poc2-contours}
\end{figure}

\subsection{PoC-3: structure recovery}\label{sec:res:poc3}
Clustering the latent is compared against clustering the shapes directly under
the geodesic metric. Three claims, in order of weight.

\paragraph{The ablation.}
Reported above (\Cref{tab:ablation}): agreement with the shape-space partition
rises from chance level to $0.781 \pm 0.049$ when the graph term is active.
This is the headline result of the study and PoC-3 is where it is measured.

\paragraph{Cluster quality.}
A high ARI alone would only show that the latent partition imitates the
reference. It does more than that: the latent partition is \emph{better
separated} than the geodesic reference at every $K$, with a silhouette of
$0.859$ against $0.632$ at $K = 4$ (a silhouette of $0$ indicates clusters no
better separated than chance, $1$ perfectly separated). The latent is not reproducing the
reference's cluster boundaries so much as finding cleaner ones.

\paragraph{Cost.}
Clustering in the latent is $40$ to $80\times$ faster than geodesic
$k$-means on the shapes, since the geodesic reference requires pairwise
great-circle distances in $320$ dimensions while the latent requires Euclidean
distances in two.

\begin{figure}[p]
  \centering
  \includegraphics[width=0.82\textwidth]{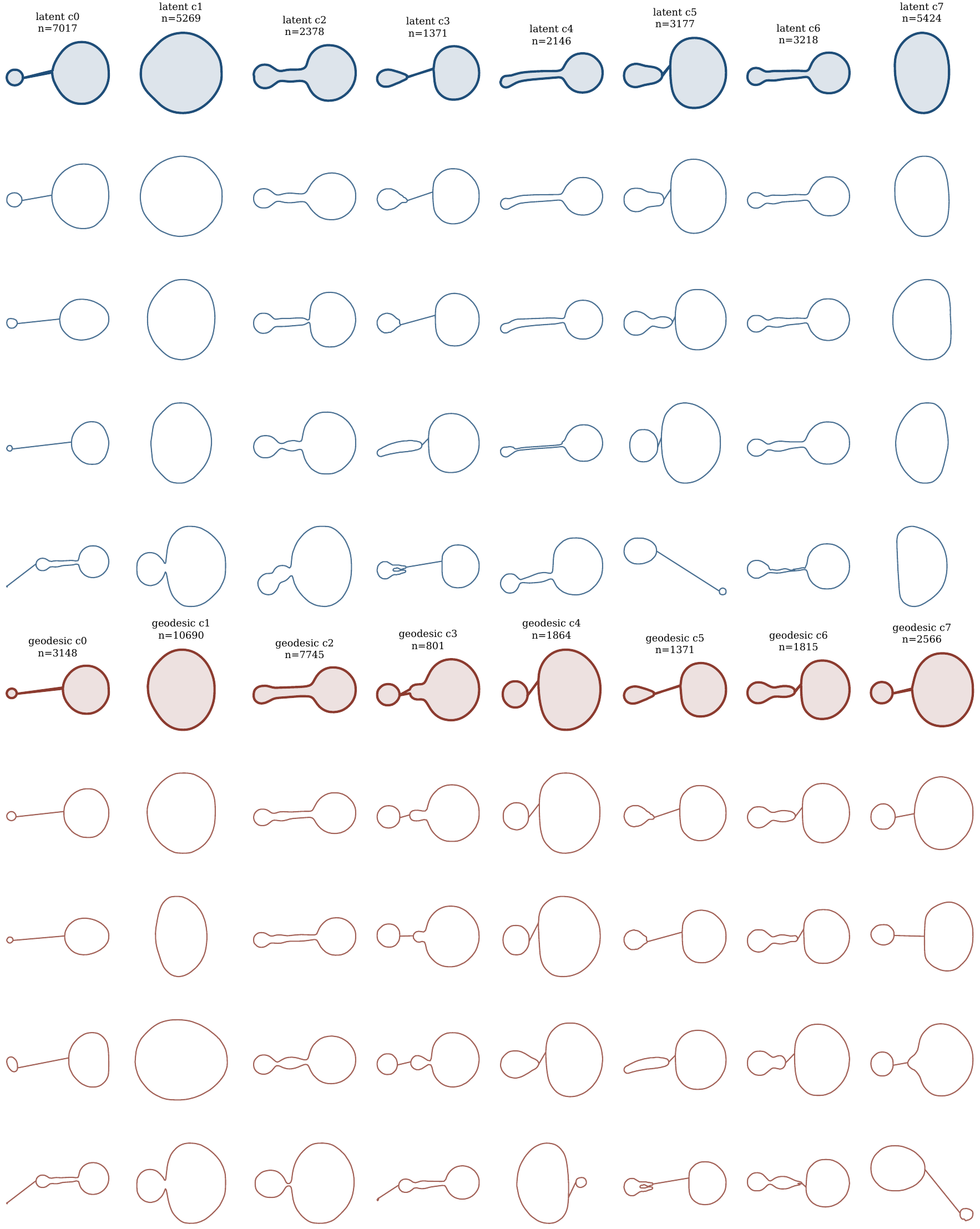}
  \caption{Decoded cluster representatives: latent clustering beside geodesic
  clustering in shape space. The last row of each block is the member farthest
  from its cluster mean, so each block shows its own worst case rather than
  only its centre.}
  \label{fig:clusters}
\end{figure}

\paragraph{Limitations of the adjusted Rand index.}
The seed-to-seed spread of ARI on this dataset is $\pm 0.04$ to $\pm 0.14$
depending on configuration. It therefore has ample resolution for the
twelve-fold ablation effect and none whatever for a few-percent difference
between two good models. ARI cannot rank models or configurations in
this study, and no ranking is claimed from it; every ARI value is quoted as
mean $\pm$ sd so that the reader can apply the same test. Differences that
would be reported as improvements under a single-seed protocol are reported
here as nulls.

\subsection{Benchmark against linear and geometric baselines}\label{sec:res:benchmark}

\shrom{} recovers the structure of shape space where the alternatives
do not, and it does so at a reconstruction cost of four percent. Against a
linear reduced-order model and against an autoencoder built to make its latent
geometrically well behaved, \shrom{} preserves $70$ to $104\%$ more of each
shape's neighbourhood and recovers a partition of shape space that the
baselines miss almost entirely (\Cref{tab:benchmark}).

The one axis on which a baseline leads is reconstruction, and it should: an
isometry-regularised autoencoder optimises for exactly that. What follows sets
out how large that lead actually is, and why it does not transfer to any task a
reduced-order model is built for.

\begin{table}[t]
  \centering
  \caption{Benchmark on the current dataset ($301{,}539$ shapes). Each baseline
  is fitted at the latent dimension it is compared against, in a single run.
  POD is deterministic; Geometric-AE is quoted as mean $\pm$ sd over three
  seeds, as is the \shrom{} clustering column.}
  \label{tab:benchmark}
  \begin{tabular}{lccc}
    \toprule
    Model & Recon.\ MSE & $k$NN overlap & ARI ($K{=}4$) \\
    \midrule
    \multicolumn{4}{l}{\emph{at $\latdim=5$}}\\
    PCA/POD               & $1.386\times10^{-4}$ & $0.514$ & --- \\
    Geometric-AE          & $\mathbf{5.51\times10^{-5}} \pm 4.2\times10^{-6}$ & $0.310 \pm 0.064$ & $0.057 \pm 0.019$ \\
    \shrom{}              & $1.217\times10^{-4}$ & $\mathbf{0.527}$ & $\mathbf{0.781 \pm 0.049}$ \\
    \midrule
    \multicolumn{4}{l}{\emph{at $\latdim=2$}}\\
    PCA/POD               & $2.821\times10^{-4}$ & $0.203$ & --- \\
    Geometric-AE          & $\mathbf{6.57\times10^{-5}} \pm 4.4\times10^{-7}$ & $0.244 \pm 0.025$ & $0.096 \pm 0.074$ \\
    \shrom{}              & $2.070\times10^{-4}$ & $\mathbf{0.498}$ & $\mathbf{0.829 \pm 0.059}$ \\
    \bottomrule
  \end{tabular}
\end{table}

Geometric-AE reconstructs better than \shrom{} at both latent dimensions, by
factors of $2.2$ to $3.2$. It preserves neighbourhood structure --- the fraction of each shape's
nearest neighbours that remain nearest in the latent --- $70\%$ worse at
$\latdim = 5$ and $104\%$ worse at $\latdim = 2$, gaps of $3.4$ and $10.1$
baseline standard deviations --- and recovers the shape-space partition far worse: ARI
$0.057 \pm 0.019$ against $0.781 \pm 0.049$ at $\latdim = 5$, and
$0.096 \pm 0.074$ against $0.829 \pm 0.059$ at $\latdim = 2$ --- separations of
$9.7$ and $5.5$ times the pooled standard deviation. At $\latdim = 5$ the
baseline is close to the chance level of zero.

The two gaps are not comparable in size, and the ratios disguise that.
On unit-norm \srvf{} vectors a mean squared error of $5.5\times10^{-5}$
corresponds to a geodesic of $0.133$~rad and one of $1.2\times10^{-4}$ to
$0.198$~rad. Against the $1.06$~rad separating two shapes drawn at random from
the corpus (\Cref{tab:scales}), both models reconstruct to within about two
percent, and the difference between them is four percent of that separation.
The clustering gap, on a scale where $0$ is chance and $1$ is exact, is $0.057$
against $0.781$. A model that reconstructs four percent better and recovers
almost none of the shape-space structure is not a better shape model.

These are not in tension. Reconstruction error measures whether the decoder
returns a shape it was given; it says nothing about whether the latent between
those points is organised. An isometry penalty on the decoder Jacobian
(\Cref{eq:isoloss}) produces a well-conditioned map and therefore good
reconstruction, without any mechanism that would arrange the latent by shape
similarity. The two objectives are close to orthogonal, and the benchmark
reports both rather than the one on which the method leads.

The margin is large enough to survive a substantial handicap: \shrom{} at
$\latdim = 2$ exceeds Geometric-AE at $\latdim = 5$ on structure overlap
($0.498$ against $0.310$) using $2.5\times$ fewer latent dimensions.

The baseline was fitted twice, from independent initialisations, as a check on
the comparison itself: reconstruction $5.51$ against $5.71 \times 10^{-5}$,
neighbourhood overlap $0.310$ against $0.338$, and clustering $0.057$ against
$0.055$. Every quantity reproduces within its seed spread, so the gaps reported
here are properties of the two methods rather than of a particular fit.

\paragraph{Transfer of the reconstruction advantage to interpolation.}
Put through the interpolation protocol at anchor step $10$, over three seeds
each, Geometric-AE reaches $0.0776 \pm 0.0020$ and the plain autoencoder
$0.0784 \pm 0.0029$, against $0.1005 \pm 0.0138$ for \shrom{}
(\Cref{tab:poc1-headtohead}). The two models optimised for fidelity are
indistinguishable from one another --- a gap of $0.0008$ against a pooled
standard deviation of $0.0025$ --- and both lead \shrom{}, which is behind the
plain autoencoder in every one of the three seeds, by $0.0221$ on average at
$\latdim = 5$. That gap clears twice the pooled standard deviation. The
ordering is identical at $\latdim = 2$ and in every seed there too, though only
the Geometric-AE margin over \shrom{} ($0.0179$) clears the bar; the plain
autoencoder's ($0.0101$) does not.

The comparison that matters here is not against Geometric-AE but against the
plain autoencoder, because the two differ only in the graph term. Removing the
term \emph{improves} interpolation error on this protocol. The isometry penalty
neither helps nor hurts: Geometric-AE and the unregularised autoencoder are
within a pooled standard deviation of each other at both latent dimensions. What the protocol responds to is reconstruction accuracy alone,
and the three models rank exactly as their reconstruction errors do.

This is the prediction of \Cref{sec:res:poc1} --- that the task rewards
reconstruction --- confirmed by a third route, after the density sweep and the
adjacent-pair test. Those three arguments together establish that the protocol saturates at the
reconstruction limit. The error stops falling as anchors are added. A reference
path that bypasses the latent entirely matches the model path. Interpolating
between adjacent frames is more accurate than the round trip through the
autoencoder at the same two frames. A benchmark whose
score is set by how well the decoder returns a shape it was given cannot
discriminate between latents on the basis of how they are organised, which is
the property \shrom{} is built to supply. We report the numbers because they
are the honest outcome of a stated protocol, and read them as a measurement of
that protocol rather than as a ranking of the models on interpolation.

The trade-off is explicit. The graph term produces the latent structure of
\Cref{tab:benchmark} --- $70$ to $104\%$ more neighbourhood
preservation and an ARI of $0.781$ against $0.057$ --- and it costs
approximately $0.022$~rad of reconstruction-driven interpolation error at
$\latdim = 5$, about two percent of the $1.06$~rad separating two shapes drawn
at random. It also costs reproducibility on this particular measure: \shrom{}'s
seed spread here is $4.8\times$ the plain autoencoder's, the one axis on which
our seed stability is worse rather than better.

\begin{table}[t]
  \centering
  \caption{Interpolation error at anchor step $10$, mean $\pm$ sample standard
  deviation over three seeds. Lower is better. The protocol saturates at the
  reconstruction limit (\Cref{sec:res:poc1}), so it orders the models by
  reconstruction accuracy rather than by latent organisation.}
  \label{tab:poc1-headtohead}
  \begin{tabular}{lcc}
    \toprule
    Model & $\latdim = 5$ & $\latdim = 2$ \\
    \midrule
    Geometric-AE      & $\mathbf{0.0776 \pm 0.0020}$ & $\mathbf{0.0794 \pm 0.0016}$ \\
    Plain autoencoder & $0.0784 \pm 0.0029$ & $0.0871 \pm 0.0059$ \\
    \shrom{}          & $0.1005 \pm 0.0138$ & $0.0972 \pm 0.0085$ \\
    \bottomrule
  \end{tabular}
\end{table}

Reconstruction is otherwise the only axis on which the baseline leads. Put through the
same downstream evaluations as \shrom{}, it recovers the shape-space partition
at close to chance (\Cref{tab:benchmark}) and predicts the latent from waveform
parameters less accurately at both dimensions ($R^2 = 0.804 \pm 0.087$ against
$0.868 \pm 0.010$ at $\latdim = 5$; $0.710 \pm 0.134$ against
$0.878 \pm 0.026$ at $\latdim = 2$). Neither regression gap clears twice the pooled standard deviation, so no
advantage is claimed there. The baseline's seed spread is nonetheless five to
nine times ours. A latent organised only by a pointwise smoothness condition is
therefore not reproducible between random initialisations, and a reduced-order
model that changes with its seed is difficult to rely on.

\paragraph{Similarity graph versus additional capacity.}
A model optimised for reconstruction alone will always win a reconstruction
contest; that is what it is for. The question this section settles is whether
winning it is sufficient, and the grid figures answer directly.
\Cref{fig:grid-ablation} decodes a regular lattice over each latent: without
the graph term four fifths of the lattice is the same near-circular oval, so
the space between the training points holds no information about shape.
\Cref{fig:latent-compare} shows the same contrast in the latent rather than in
decoded shapes.

This is what the graph term is for, and it cannot be bought with capacity or
with a smoother decoder. The reconstruction objective constrains
$\dec \circ \enc$ at the observed points and is indifferent everywhere else; a
Jacobian penalty adds a pointwise smoothness condition but still says nothing
about which shapes should be near which.

That distinction is what makes the interpolation task of \Cref{sec:res:poc1}
possible at all. A filmed sequence is sparse in time, but the corpus is dense
in shape: a state the camera missed between two frames is not a novel
morphology, it resembles shapes that other actuation settings produced and that
were recorded. Supplying it means recognising that resemblance, which requires
a term that encodes which shapes are alike. No amount of reconstruction
accuracy on the frames that \emph{were} captured provides that. Only a term that encodes the neighbourhood structure of the data can make
latent \emph{position} correspond to morphology. Every downstream use of a
reduced-order model reads that position: sampling a grid, interpolating a
trajectory, inverting from a design parameter. None of them reads the
reconstruction of an already-observed point.

\paragraph{Grid occupancy, where the baseline leads.}
By grid occupancy, Geometric-AE's latent is \emph{less} sparse than ours:
$77.8\%$ of its grid lies far from any training point, against $98.1\%$ for
\shrom. The isometry penalty spreads points evenly across the latent, whereas
the graph term deliberately clusters them and leaves the space between clusters
empty. Even spreading and meaningful structure are different properties, and a
reader may reasonably prefer the former for some purposes; the number is
reported rather than omitted.

\subsection{Variants}\label{sec:res:variants}
Two architectural variants were tested and neither is adopted.

Rarity-weighted reconstruction, which up-weights infrequent morphologies in the
reconstruction term, improves nothing measurably on any task. A wider decoder
is nominally better at interpolation but does not reach significance, and costs
four times the decoder parameters; the narrower network is therefore the
headline on parsimony. Both are reported here as null results, in the same
terms as the positive ones.

\subsection{Sensitivity to the neighbour-graph metric}\label{sec:res:metric}

The regulariser needs a distance from which to build its neighbour graph, and
\srvf{}s lie on a sphere, so the choice of Euclidean distance
(\Cref{sec:setup:metric}) invites the objection that a spherical metric would
be more appropriate. \Cref{tab:metric} tests it across five choices, two latent
dimensions and three seeds.

\begin{table}[t]
  \centering
  \caption{Five neighbour-graph metrics, two latent dimensions, three seeds
  each. Every comparison against Euclidean is non-significant under the
  two-pooled-sd rule of \Cref{sec:setup:stats}: $0$ of $52$.}
  \label{tab:metric}
  \resizebox{\textwidth}{!}{
  \begin{tabular}{lcccc}
    \toprule
    Graph metric & PoC-1 geodesic err. & PoC-2 $R^2$ & ARI $K{=}4$ & edge Jaccard vs.\ Euclid. \\
    \midrule
    Euclidean            & $0.107 \pm 0.008$ & $0.878 \pm 0.026$ & $0.829 \pm 0.059$ & $1.000000$ \\
    cosine               & $0.119 \pm 0.014$ & $0.884 \pm 0.007$ & $0.844 \pm 0.028$ & $1.000000$ \\
    exact geodesic       & $0.113 \pm 0.015$ & $0.886 \pm 0.019$ & $0.875 \pm 0.020$ & $1.000000$ \\
    Minkowski $p{=}3$    & $0.113 \pm 0.003$ & $0.884 \pm 0.025$ & $0.818 \pm 0.034$ & $0.502$ \\
    Minkowski $p{=}1$    & $0.111 \pm 0.015$ & $0.886 \pm 0.008$ & $0.839 \pm 0.036$ & $0.389$ \\
    \bottomrule
  \end{tabular}}
\end{table}

Not one of the $52$ comparisons against Euclidean reaches
significance, and none approaches it: the largest ratio of gap to threshold is
$0.85$, where $1.0$ is the bar.

Two readings follow, of unequal interest.

The first is expected. Euclidean, cosine and the exact geodesic build the
\emph{identical} neighbour graph --- each is strictly monotone in the
great-circle angle, so the edge sets coincide exactly and only the fuzzy edge
weights differ, by around $10^{-5}$. That these three agree confirms what
\Cref{sec:method:graph} argues analytically: the regulariser consumes the
graph, not the distance values that produced it.

The second is not expected. The Minkowski arms build genuinely \emph{different}
graphs, sharing only $39\%$ and $50\%$ of their edges with Euclidean and
ranking neighbours differently (rank correlation $0.878$ and $0.916$ against
the geodesic, against $1.000000$ for the other three). They were predicted to
perform worse. They do not. The supportable claim is therefore stronger than
the monotonicity argument alone would give: the downstream tasks are
insensitive to changing half the edges of the neighbour graph.

Euclidean is retained on parsimony rather than superiority. Two cautions
belong with this result. Three seeds establishes the absence of a large effect
and not the absence of an effect, and here the seed spread is comparable to
every gap under test. And the nominally best clustering figure in
\Cref{tab:metric} belongs to the geodesic arm; at $0.55$ of the significance
threshold it is seed noise, and is not a result.

\paragraph{Interpretation of the per-metric grids.}
Latent grids for the individual metric arms are single-seed illustrations and
should be read as such. On the Euclidean and cosine pair, the count of
self-intersecting grid nodes across the three seeds was $13/33/12$ and
$2/23/18$ respectively --- a difference of $5$ against a pooled spread of
$22.8$ --- yet the first seed alone would read as a sixfold improvement. The
verdict on this question comes from \Cref{tab:metric}, not from the figures.

\subsection{Regularisation strength and the locality--globality trade-off}\label{sec:res:umap}

The kernel of \Cref{eq:qij} has an effective width, set through
$\texttt{min\_dist}$, controlling how firmly non-neighbours are pushed apart.
Unlike the graph metric, this parameter does have an effect --- and the three
proofs of concept want opposite things from it, so no single setting serves all
three.

\begin{table}[t]
  \centering
  \caption{Four regularisation strengths, three seeds each. The kernel
  parameter $a$ spans an $8\times$ range; similarity at unit latent distance
  spans $0.35$ to $0.81$.}
  \label{tab:umap}
  \resizebox{\textwidth}{!}{
  \begin{tabular}{lccccc}
    \toprule
    Config & \texttt{min\_dist} & PoC-1 @120 & PoC-2 $R^2$ & ARI $K{=}4$ & anisotropy \\
    \midrule
    continuous  & $0.8$  & $\mathbf{0.114 \pm 0.007}$ & $0.870 \pm 0.023$ & $0.743 \pm 0.041$ & $4.5$ \\
    mid         & $0.3$  & $0.122 \pm 0.005$ & $0.853 \pm 0.019$ & $0.755 \pm 0.033$ & $13.1$ \\
    default     & $0.1$  & $0.123 \pm 0.003$ & $0.864 \pm 0.013$ & $\mathbf{0.851 \pm 0.014}$ & $41.1$ \\
    manuscript  & $0.01$ & $0.125 \pm 0.002$ & $0.867 \pm 0.008$ & $0.846 \pm 0.021$ & $36.9$ \\
    \bottomrule
  \end{tabular}}
\end{table}

Three readings of \Cref{tab:umap}.

Interpolation error falls \emph{monotonically} with $\texttt{min\_dist}$, from
$0.125$ to $0.114$, a reduction of $9.1\%$, and the ordering reproduces across
seeds. The best-versus-baseline gap is nonetheless only $1.5$ pooled standard
deviations. We therefore report the trend rather than the gap: a monotone
ordering across four configurations has a probability of about $4\%$ under the
null, which is evidence, while the individual difference is not.

Clustering runs the other way, leading at low $\texttt{min\_dist}$. This is the
locality--globality trade-off the parameter is expected to produce: a tightly
packed embedding separates groups and a loosely packed one connects them, and
the two tasks reward opposite ends of that scale. Regression is insensitive to
the choice.

\begin{figure}[t]
  \centering
  \begin{subfigure}{0.48\textwidth}
    \includegraphics[width=\textwidth]{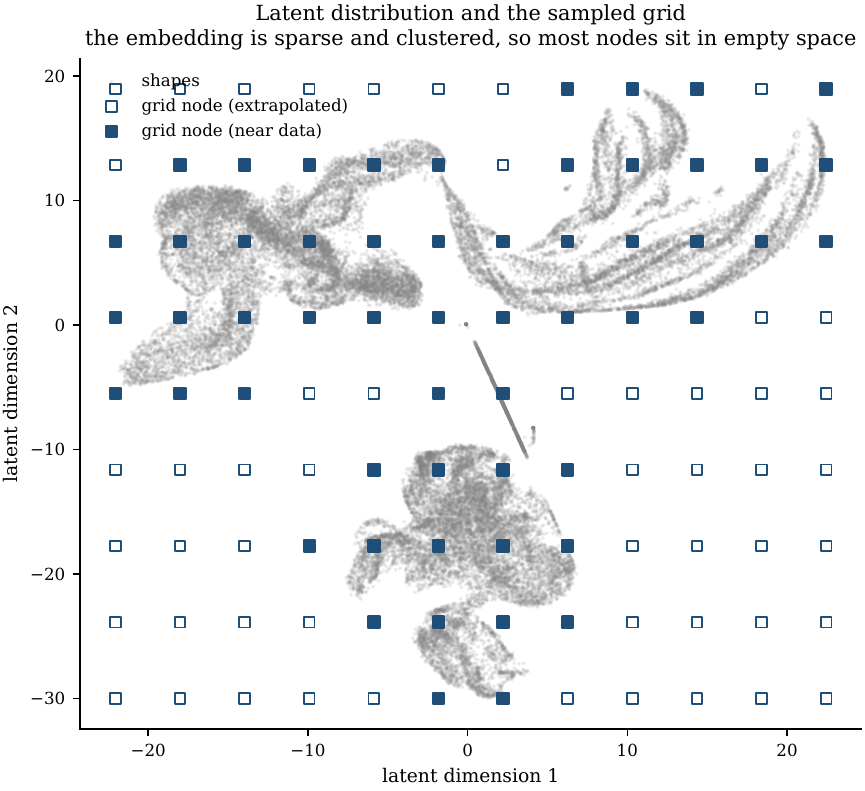}
    \caption{latent cloud and sampled grid}
  \end{subfigure}\hfill
  \begin{subfigure}{0.48\textwidth}
    \includegraphics[width=\textwidth]{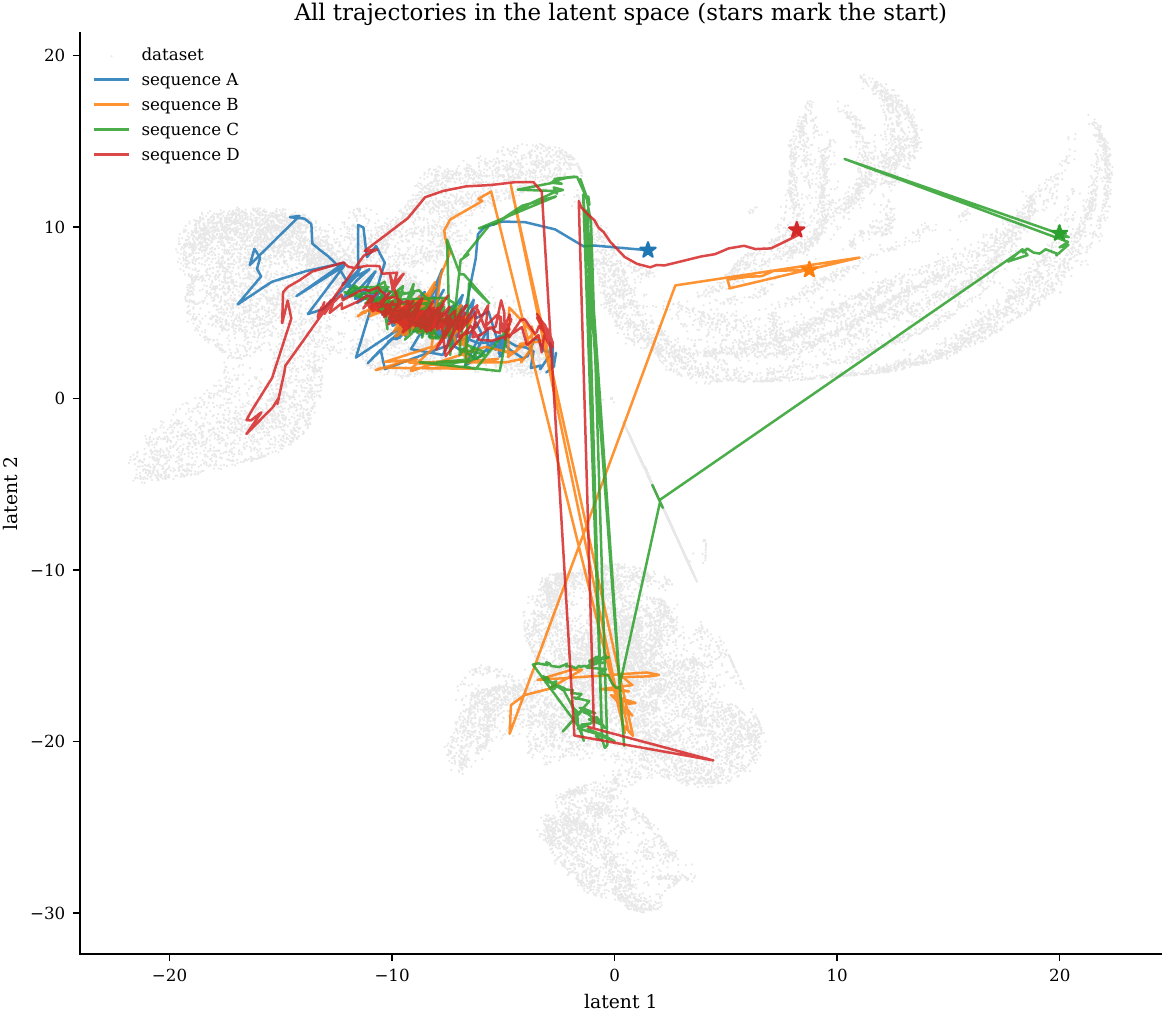}
    \caption{the four sequences through the latent}
  \end{subfigure}
  \caption{The continuous configuration ($\texttt{min\_dist} = 0.8$): the latent
  cloud with the sampled grid, and the four filmed sequences drawn through it.
  The pinch-off behaviour of all three configurations is in
  \Cref{fig:pinchoff}.}
  \label{fig:pinchoff-continuous}
\end{figure}

\paragraph{Effect of a more continuous latent.}
\Cref{fig:pinchoff-continuous} repeats \Cref{fig:pinchoff} --- the same frames
and sub-steps --- under the most continuous setting. Intervals within a
topology class are visibly smoother. The pinch-off still collapses. Quantified,
the latent jump falls from $513\times$ a normal step to $60\times$, an
$8.6\times$ improvement that nonetheless leaves a discontinuity of two orders
of magnitude.

This is the cleanest evidence that the separation is a property of the
representation rather than of the embedding. The regulariser is the only
mechanism available for connecting the latent, it compresses the gap by nearly
an order of magnitude, and it cannot remove it.

\section{Discussion}\label{sec:discussion}

The results establish that the graph term determines whether latent position
corresponds to morphology. This section asks what follows from that: why the
term is necessary, where the method stops working, which alternatives were
considered and rejected, and what the negative results indicate. No new
measurements are introduced.

\subsection{The role of the manifold-informed loss}\label{sec:disc:why}

A corpus dense in shape is only useful if the model knows which shapes are
alike, and the results say that knowledge has to be asked for explicitly. The
graph term affects one quantity, narrowly. It does not improve reconstruction
or regression accuracy. It determines whether clustering the latent recovers the
structure that clustering the shapes themselves would find, and it does so by
more than thirteen pooled standard deviations.

The reason is that a reconstruction objective is indifferent to how the latent
is arranged. It constrains the composition $\dec \circ \enc$ at the training
points and says nothing about the space between them. A model can satisfy it
completely while filling the gaps between its data with copies of the mean,
which is what the plain autoencoder does: four fifths of its decoded grid is
the same oval, and its latent partition is indistinguishable from chance.

The graph term supplies the missing constraint. It does not make
reconstruction better --- it makes \emph{position} in the latent carry
information. For a model that will only ever be evaluated at observed points,
this is a distinction without a difference. For a model intended to be
\emph{sampled} --- interpolated through, gridded over, inverted from a design
parameter --- it is the property that decides whether the output is meaningful,
and it is invisible to reconstruction error.

\subsection{The central limitation: topology change}\label{sec:disc:limitation}

The degenerate connector of \Cref{sec:method:bridge} makes a multi-component
interface representable, and the two topology classes it distinguishes end up
far apart in the latent. The measurements make the gap explicit: the classes sit
thirty-three times further apart than the typical within-class spacing,
consecutive frames move thirty-six times further across a change than within
one, and at pinch-off the latent jumps by up to two orders of magnitude more
than a normal step. Whether the connector \emph{causes} that gap or merely
makes it visible is settled below.

This is a representational failure, not an interpolation failure, and
the distinction matters for anyone trying to fix it. The fitted paths through
these intervals are well behaved --- monotone between their endpoints,
overshooting nowhere, verified numerically. The problem is that the latent
region they traverse contains no valid intermediate shapes. No interpolation
scheme can produce intermediates that the representation does not contain, so
no better path-fitting method will help.

The regularisation experiments sharpen this. Increasing
$\texttt{min\_dist}$ compresses the pinch-off discontinuity by nearly an order
of magnitude and cannot remove it, and the setting that compresses it most
produces the least plausible path through it. The separation is a property of
the shape representation, and the embedding can only attenuate it.

Three further measurements make the same point from different directions.
First, the separation in the latent neighbour graph is absolute: across four
regularisation strengths and four latent dimensions, no connected component
contains both a single-component and a bridged shape
(\Cref{tab:connectivity}). Raising $\texttt{min\_dist}$ reconnects fragments
\emph{within} a topology class and never across one, and raising the latent
dimension to the estimated intrinsic dimension of the data
(\Cref{sec:res:dimension}) does not merge them either.

Second, a variant trained as one model per topology class, which removes the
divide by making it unrepresentable, interpolates roughly seven times worse
than any other configuration: a trajectory crosses the divide partway through,
so no such model ever sees a whole sequence.

Third, and most directly, the plain autoencoder does not separate the two
families at all. Its largest component holds $95.3\%$ of the points and mixes
the single-component and bridged populations freely. That contrast locates the
divide precisely. It is not produced by the connector, nor by the embedding
dimension, nor by the kernel: it appears exactly when the loss is told which
shapes are alike, and it is absent when the loss is indifferent to shape
similarity. The plain autoencoder merges the two families only because position
in its latent carries no shape meaning --- the same defect the ablation of
\Cref{sec:res:ablation} measures on clustering.

The divide is therefore not an artefact to be engineered away. It is the
regulariser reporting a genuine separation between two shape families that a
breakup event creates, and a representation that hides it is worse, not better.
The limitation is confined to interpolation across that event. Everywhere else
in the latent, position continues to correspond to morphology.

\subsection{Alternatives considered}\label{sec:disc:alternatives}

Four alternatives would address the topology limitation more fundamentally, and
each was considered and set aside for a stated reason. \emph{Elastic shape
graphs} model branching structure directly. Implementations exist for specific
domains --- retinal vessel networks, for instance --- but we found none written
for planar fluid interfaces, and adapting a domain-specific one is a project in
itself rather than a component to be reused. The \emph{extended SRVF} with
branch birth and death is designed for three-dimensional tree structures and
does not transfer cleanly to planar interfaces. \emph{Varifolds} handle
topology change naturally but abandon the elastic metric, and with it the
reparametrisation invariance that the interpolation and clustering results
depend on. \emph{Wasserstein--Fisher--Rao} distances accommodate mass creation
and destruction but require leaving the \srvf{} parametrisation entirely.

We are not aware of published work that quantifies connector distortion inside
an elastic metric. The measurements in \Cref{sec:res:poc1} are that
quantification, offered as a characterisation of the problem rather than a
solution to it.

\subsection{Negative results}\label{sec:disc:nulls}

Several things did not work, and they are collected here rather than
distributed thinly through the results.

Rarity-weighted reconstruction improves nothing. A wider decoder is nominally
better at interpolation but does not reach significance at four times the
parameter count. The adjusted Rand index has a seed spread that makes it
incapable of ranking two good models, whatever its resolution for the ablation.
The regularisation strength shows a monotone trend across four configurations
that does not clear two pooled standard deviations at any single comparison.
And an isometry-regularised baseline reconstructs substantially better than we
do.

The most instructive null is that the plain autoencoder wins the
interpolation task. It does so at every anchor density, by a margin that does
not close as anchors are added. Rather than report that and move on, we traced
it: the error saturates well above zero however densely the trajectory is
sampled, and joining two consecutive frames by a straight line --- the exact
interpolant, with nothing left to tune --- costs no measurable accuracy at all.
The quantity PoC-1 measures is therefore reconstruction fidelity, which is
precisely what the manifold term trades away. A proof of concept can be
well-posed and still measure something other than what it was designed to
test, and finding that out required three experiments that a positive result
would never have prompted.

The neighbour-graph metric deserves more than a clause. Five metrics
were tested across two latent dimensions and three seeds, and not one of the
fifty-two comparisons against Euclidean distance reached significance. This
includes metrics that build a demonstrably different graph, sharing under half
their edges with the Euclidean one. The practical consequence is worth stating
plainly: a practitioner adopting this method need not agonise over the distance
function, provided the graph is built over a shape space at all. Establishing
that cost five trained configurations, and it is the kind of result that is
rarely published and frequently re-derived.

\subsection{Suitability of an isometry penalty}\label{sec:disc:isometry}

The benchmark result --- a baseline that reconstructs several times better and
recovers shape structure several times worse --- is not a concession but a
consequence, and two published lines of work explain it.

\paragraph{Local versus global structure.}
The penalty $\lVert J^{\top}J - cI\rVert_F^2$ constrains the decoder Jacobian
\emph{at a point}. A map may be a perfect local isometry everywhere and still
tear the manifold or fold distant regions onto one another, because a pointwise
condition on the Jacobian cannot see that two shapes far apart in the data have
been mapped together. The graph term is the opposite kind of constraint: an
explicit, dataset-wide statement about which points must remain near which.

\paragraph{Isometry with respect to the wrong geometry.}
The penalty preserves ambient Euclidean geometry of the coordinate vector, but
\srvf{}s lie on a sphere where the meaningful distance is arc length.
\citet{jang2023geometrically} show that applying vector-space regularisation to
data on a curved space can degrade autoencoder performance outright; our result
is an instance of exactly that. \citet{osipov2026geometry} report the same
effect in the closest published setting to ours, finding that near-isometry
regularisation of the decoder Jacobian made downstream latent-dynamics training
\emph{harder} in an autoencoder reduced-order model, even where it improved
local decoder smoothness.

Neither observation makes the isometry penalty a poor method. It makes it a
method optimised for a property --- decoder conditioning --- that is close to
independent of the property a generative shape model needs.

\subsection{Latent compression relative to the data dimension}\label{sec:disc:dimension}

A five-dimensional latent uses about one and a fifth of its dimensions and
concentrates almost all of its variance in two. This is a property of the
trained decoder rather than of the data: the shape corpus itself is not
two-dimensional, and the estimators of \Cref{sec:res:dimension} place its
intrinsic dimension between two and twelve depending on the scale examined. The
representation therefore compresses well below the dimension of its input. It
can do so because the directions it discards do not affect the quantities the
downstream tasks require, which is what the dimension comparison tests
directly.

One observation resists an easy explanation. A wider decoder concentrates the
latent \emph{further} rather than using more of it, which is the opposite of
what additional capacity might be expected to do. We report it without a
mechanism.

\subsection{Limits of a purely geometric representation}\label{sec:disc:physics}

Nothing in the model knows any fluid mechanics. A state produced between two
observations is whatever the latent geometry places there, constrained only by
resemblance to shapes the corpus contains. That the results hold without any physical
constraint is itself informative, since it shows how much geometric resemblance
alone carries. It also bounds them. Physical constraints identify where the
reconstruction-limited floor of \Cref{sec:res:poc1} might be beaten, which
further tuning of the regulariser cannot achieve.

Two routes are open, at different cost. A term in the training objective could
penalise physically inadmissible intermediate states: volume that is not
conserved between break-up events, or curvature evolution inconsistent with
surface tension. This shapes the latent itself, so that the space between
observations contains only realisable morphologies, and is the more fundamental
of the two. Alternatively the path fitting could be constrained. The schemes
used here are purely geometric --- splines and geodesics that know nothing of
the dynamics --- and even a scalar conservation law imposed along the trajectory
would tailor the traversal between observations without retraining anything.
The second is much the cheaper experiment and is where we would begin.

Neither is attempted here, and the results should be read as what a
similarity-organised latent achieves on its own.

\subsection{Generality}\label{sec:disc:generality}

The recipe --- a metric shape space, a neighbour graph over it, an autoencoder
regularised by that graph --- makes no reference to droplets. It should apply
wherever shapes admit a metric representation and the latent is meant to be
sampled rather than only evaluated: cell morphology, airfoil families, bubble
and bridge dynamics, and free-surface flows generally. That transfer is
untested here and is asserted only as a design claim.

One qualification follows from our own results. The recipe should not be stated
as requiring a neighbour graph built \emph{in the shape metric}: the ablation
of \Cref{sec:res:metric} shows that metrics sharing fewer than half their edges
give indistinguishable outcomes. What matters is that the graph is computed
over a space in which proximity means shape similarity, not which distance
function computes it. That formulation is weaker, better supported by the
evidence, and considerably easier to adopt.

\section{Conclusions}\label{sec:conclusions}

An engineer with a high-speed recording of a break-up has, at best, a dozen
frames of an event lasting tens of microseconds, and no way to obtain what
happened between them short of repeating the experiment. What that engineer
does have, if the campaign swept the device's operating range, is a large
collection of other droplets --- among them shapes closely resembling the ones
the recording missed.

\shrom{} makes that collection usable. Representing interfaces by their
square-root velocity functions gives a metric in which resemblance is
meaningful. Building a neighbour graph over that space, and regularising an
autoencoder with it, produces a latent in which position corresponds to
morphology. The decoder maps any latent position back to an interface. A state
absent from one sequence is then recoverable from the corpus that contains its
likeness.

The composition achieves what none of its parts achieves alone. Removing
the graph term from an otherwise identical model leaves reconstruction
unchanged but destroys the correspondence between latent position and shape.
The latent then becomes no more informative about shape structure than a random
partition. The graph term therefore does not improve reconstruction. It
establishes a correspondence between latent position and morphology at
coordinates where no training shape lies. A model with this property can be
sampled at unobserved coordinates; a model without it can only be evaluated at
observed ones. Reconstruction error does not measure the property, so no
autoencoder acquires it by default.

The learned representation uses two dimensions, although the shape corpus
itself is of higher intrinsic dimension. The effective dimensionality of the
trained decoder indicated this before the downstream tasks confirmed it. The
entire morphology family can therefore be drawn on one plane and read directly,
giving a reduced-order model whose latent is small enough to inspect rather than
only to query.

Three things were not shown. Interpolation across a change of interface
topology does not work, and the measurements locate the cause in the
representation rather than in the interpolation. No path-fitting scheme can
supply intermediates that the representation does not contain. The separation
in the latent neighbour graph is absolute: across four regularisation strengths
and four latent dimensions, no connected component holds both a
single-component and a post-break-up shape. Training a separate model per
topology class removes the divide by making it unrepresentable, and interpolates
substantially worse, because a trajectory crosses the divide partway through.
An unregularised autoencoder, by contrast, places the two families in one
component and mixes them freely. That contrast identifies the cause. It is not
the connector, and not the embedding dimension, but the requirement that latent
position track shape similarity. The separation is a real discontinuity in
shape space that the regulariser reports faithfully. Several plausible design choices turned out
not to matter, among them the metric used to build the neighbour graph, which
we had expected to be consequential. And a baseline optimising a different
notion of latent geometry reconstructs better than \shrom{} does, on an axis
that we argue is the wrong one but do not dismiss.

The interpolation task deserves a sentence of its own, because chasing its
result taught us what it measures. A plain autoencoder predicts withheld frames
more accurately than \shrom{} does, and the gap does not close as the
trajectory is sampled more densely. Joining two consecutive frames by a
straight line --- an exact interpolant, with nothing left to fit --- costs no
measurable accuracy, which locates the entire residual in the reconstruction
rather than in the path. What the task rewards is therefore fidelity, and
fidelity is what the manifold term spends. The proof of concept still shows
what it was built to show, that the latent yields plausible droplets at
arbitrary intermediate times; it does not show what its headline number
appeared to promise.

The most direct route past the topology limitation is a representation that
admits component birth and death natively --- elastic shape graphs, or a
measure-valued formulation --- at the cost of the reparametrisation guarantees
the present results rely on. Nearer at hand, the component count could be
encoded as an explicit channel rather than left implicit in the connector, and
the outlier filter could adapt its threshold per cluster rather than removing a
fixed fraction. Whether the method transfers to other interface families ---
bubbles, films, free surfaces --- is untested here and is the question we would
ask first.

\section*{Data availability}
The interface contours analysed in this work are publicly available at
\url{https://doi.org/10.5281/zenodo.14275300} \citep{hashemi2024contours}. The
archive contains the raw extracted contours from which every dataset in this
paper is derived; the construction is deterministic given the configuration
recorded in the exported file (\Cref{sec:app:repro}). The code implementing
\shrom{} is available from the corresponding author on reasonable request.

\appendix

\section{Supplementary material}\label{sec:appendix}

\subsection{Cluster atlas}\label{sec:app:atlas}
Per-cluster detail supporting the stage-2 filtering decision of
\Cref{sec:setup:cleaning}. No cluster is predominantly defective, which is why
filtering removes individual shapes by their distance from a cluster mean
rather than discarding whole clusters.

\begin{figure}[h]
  \centering
  \includegraphics[width=0.7\textwidth]{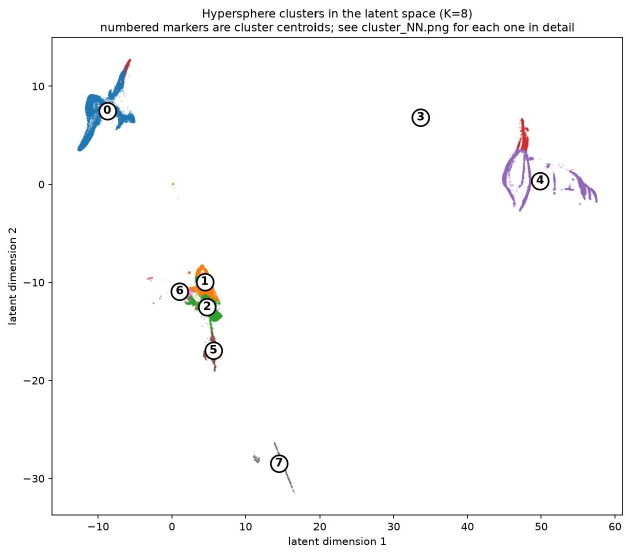}
  \caption{Hypersphere clusters in the latent, numbered, with representative
  members of each.}
  \label{fig:atlas}
\end{figure}

\subsection{Coverage of the sampled parameter space}\label{sec:app:coverage}
Parameter entropy efficiency ranges from $0.79$ to $1.00$ across the eight
waveform parameters, and shape-space occupancy is $0.843$. Two clusters contain
only eight members each. These thin regions are rare morphologies rather than
gaps in the sampling, which matters because a reduced-order model asked to
generate in a thin region is extrapolating even though the parameter
combination is nominally covered.

\begin{figure}[h]
  \centering
  \includegraphics[width=0.85\textwidth]{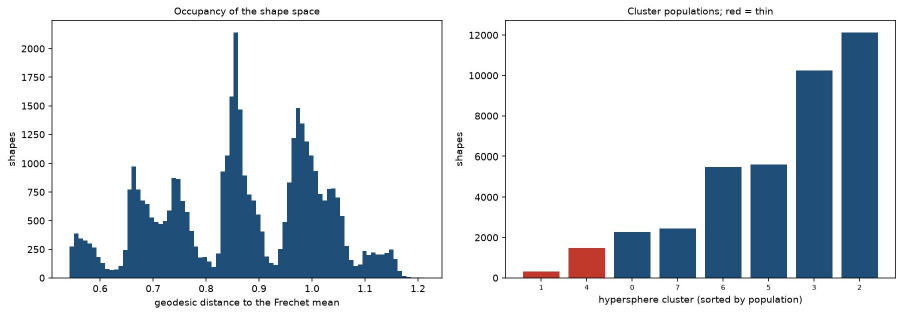}
  \caption{Shape-space occupancy against the sampled parameter space.}
  \label{fig:coverage}
\end{figure}

\subsection{Latent trajectories}\label{sec:app:paths}
The four filmed sequences drawn through the latent. Each is a path rather than
a cloud, and the topology changes are visible as jumps; \Cref{sec:res:poc1}
quantifies their size.

\begin{figure}[h]
  \centering
  \includegraphics[width=0.7\textwidth]{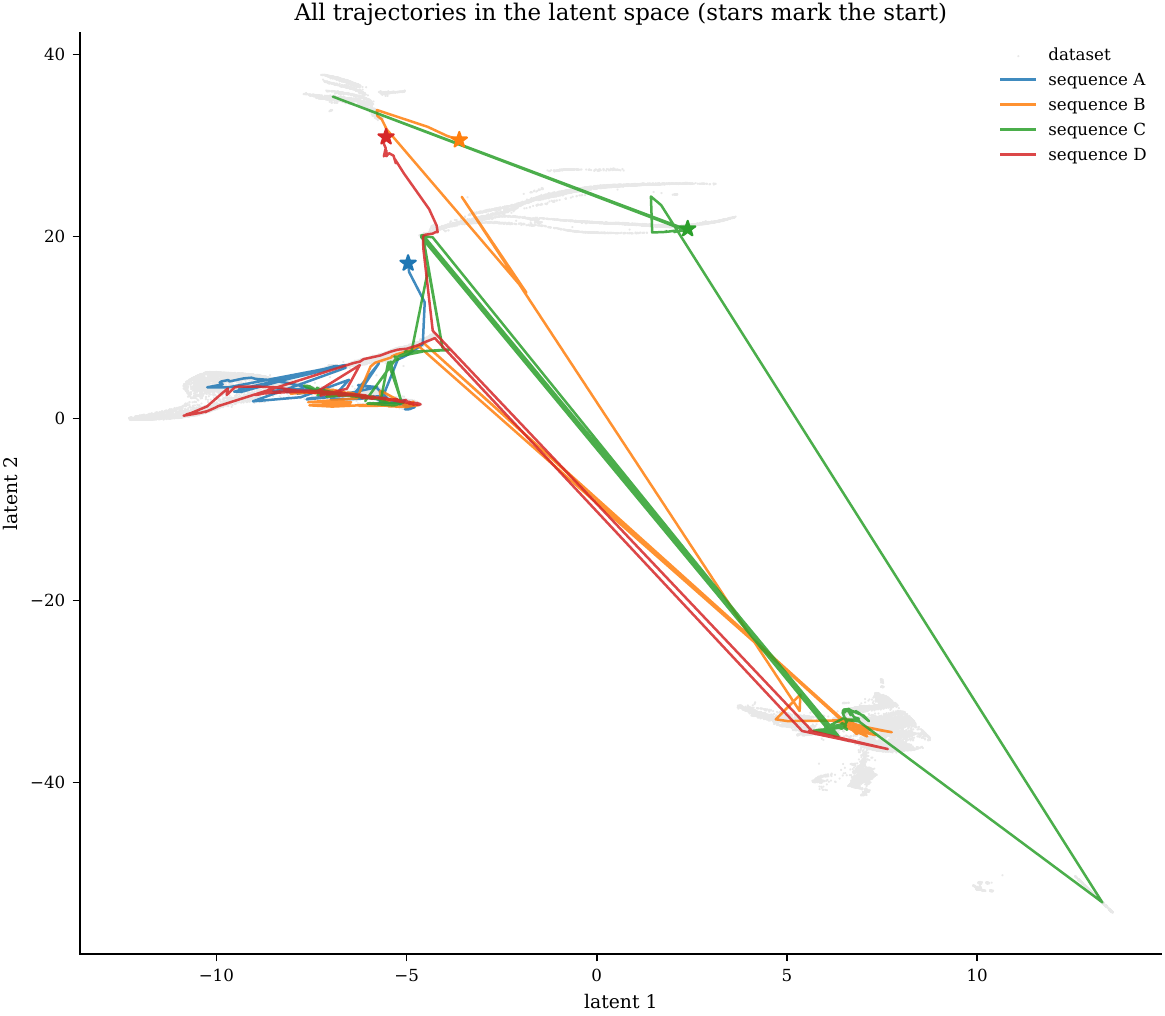}
  \caption{The four filmed sequences drawn through the latent, in acquisition
  order.}
  \label{fig:paths}
\end{figure}

\subsection{Latent connectivity}\label{sec:app:connectivity}

\Cref{sec:disc:limitation} argues that the topology divide is a property of the
representation. This is the measurement behind that claim. A $k$-nearest-neighbour
graph ($k=15$) is built over the latent and its connected components counted: a
fragmented latent has many components, a traversable one has few.

\begin{table}[h]
  \centering
  \caption{Connected components of the latent neighbour graph, $20{,}000$
  points, with each component labelled by the interface topology of the shapes
  it contains. The sample is $61.3\%$ single-component and $38.7\%$ bridged.
  Every component of every graph-regularised latent is topologically pure; the
  plain autoencoder is the only configuration whose largest component mixes the
  two.}
  \label{tab:connectivity}
  \begin{tabular}{lcccc}
    \toprule
    Configuration & Components & Largest & Mixed & Mean $k$NN distance \\
    \midrule
    plain autoencoder            & $3$  & $95.3\%$ & $1$ & $0.0176$ \\
    $\texttt{min\_dist}=0.01$    & $24$ & $35.5\%$ & $0$ & $0.0044$ \\
    $\texttt{min\_dist}=0.1$     & $9$  & $35.5\%$ & $0$ & $0.0058$ \\
    $\texttt{min\_dist}=0.3$     & $5$  & $61.3\%$ & $0$ & $0.0108$ \\
    $\texttt{min\_dist}=0.8$     & $4$  & $61.3\%$ & $0$ & $0.0195$ \\
    \midrule
    $\latdim=5$                  & $10$ & $35.5\%$ & $0$ & $0.0062$ \\
    $\latdim=8$                  & $12$ & $35.5\%$ & $0$ & $0.0129$ \\
    $\latdim=10$                 & $9$  & $35.5\%$ & $0$ & $0.0067$ \\
    \bottomrule
  \end{tabular}
\end{table}

Three readings. The manuscript setting has the \emph{smallest} neighbour
spacing and the \emph{most} components, so it is tightly clumped into islands
rather than merely spread out, which is what makes it good for clustering and
poor for interpolation.

The $61.3\%$ ceiling is exactly the single-component share of the sample. At
$\texttt{min\_dist} = 0.3$ and $0.8$ the largest component is the whole
single-component class and nothing else; at the smaller settings that class
itself fragments, and the largest component falls to $35.5\%$, or $58\%$ of
the class. Raising $\texttt{min\_dist}$ therefore reconnects fragments
\emph{within} a topology class and never across one.

The separation is absolute rather than partial. Across four
$\texttt{min\_dist}$ settings and four latent dimensions, no component of any
graph-regularised latent contains both a single-component and a bridged shape.
The plain autoencoder is the exception that identifies the cause: without the
graph term its largest component holds $95.3\%$ of the points and mixes
$12{,}261$ single-component shapes with $6{,}806$ bridged ones. A latent whose
position carries no shape meaning places the two families together; a latent
organised by shape similarity keeps them apart, because in the shape space they
\emph{are} apart. The divide is therefore a property the regulariser reports
faithfully, not one the embedding introduces.

\subsection{Latent geometry diagnostics}\label{sec:app:geometry}

The diagnostics defined in \Cref{sec:method:diagnostics} were also computed
across the graph-metric arms. They are reported here rather than in the main
text because they do not discriminate. Of eighteen pairwise comparisons, one
reaches significance under the two-pooled-standard-deviation rule, which is the
rate chance alone would produce at that threshold.

\begin{table}[h]
  \centering
  \caption{Latent geometry across four graph-metric arms, three seeds,
  $\latdim=2$. \texttt{step\_cv} is the coefficient of variation of the
  geodesic step between adjacent decoded lattice nodes, so zero would be a
  scaled isometry; \texttt{valid\_frac} is the fraction of lattice nodes
  decoding to a simple polygon; \texttt{spearman} is the rank correlation
  between latent and shape-space distance.}
  \label{tab:geometry}
  \begin{tabular}{lccc}
    \toprule
    Arm & \texttt{step\_cv} & \texttt{valid\_frac} & \texttt{spearman} \\
    \midrule
    Euclidean, $\texttt{min\_dist}=0.01$ & $1.065 \pm 0.153$ & $0.761 \pm 0.146$ & $0.734 \pm 0.224$ \\
    Euclidean, $\texttt{min\_dist}=0.8$  & $1.261 \pm 0.108$ & $0.778 \pm 0.069$ & $0.744 \pm 0.036$ \\
    cosine, $\texttt{min\_dist}=0.01$    & $1.080 \pm 0.132$ & $0.823 \pm 0.135$ & $0.714 \pm 0.109$ \\
    cosine, $\texttt{min\_dist}=0.8$     & $1.293 \pm 0.062$ & $0.790 \pm 0.160$ & $0.729 \pm 0.039$ \\
    \bottomrule
  \end{tabular}
\end{table}

The seed spread exceeds every between-arm difference. We report the table
because a null is worth recording --- these are the natural diagnostics to
reach for, and the finding is that they lack the resolution to rank
configurations on this data --- but no conclusion in the main text rests on
them. The connectivity measurement of \Cref{sec:app:connectivity}, by contrast,
separates the same configurations by a factor of six and is used in
\Cref{sec:disc:limitation}.

\subsection{Regularisation-strength study}\label{sec:app:umap}

The latent grids below show what $\texttt{min\_dist}$ does to the space itself.
The continuous setting fills it evenly; the manuscript setting concentrates it
onto one strongly anisotropic axis. Both decode to plausible shapes throughout,
so this is a difference in organisation rather than in validity, and it is the
same difference \Cref{tab:connectivity} counts.

\begin{figure}[h]
  \centering
  \begin{subfigure}{0.48\textwidth}
    \includegraphics[width=\textwidth]{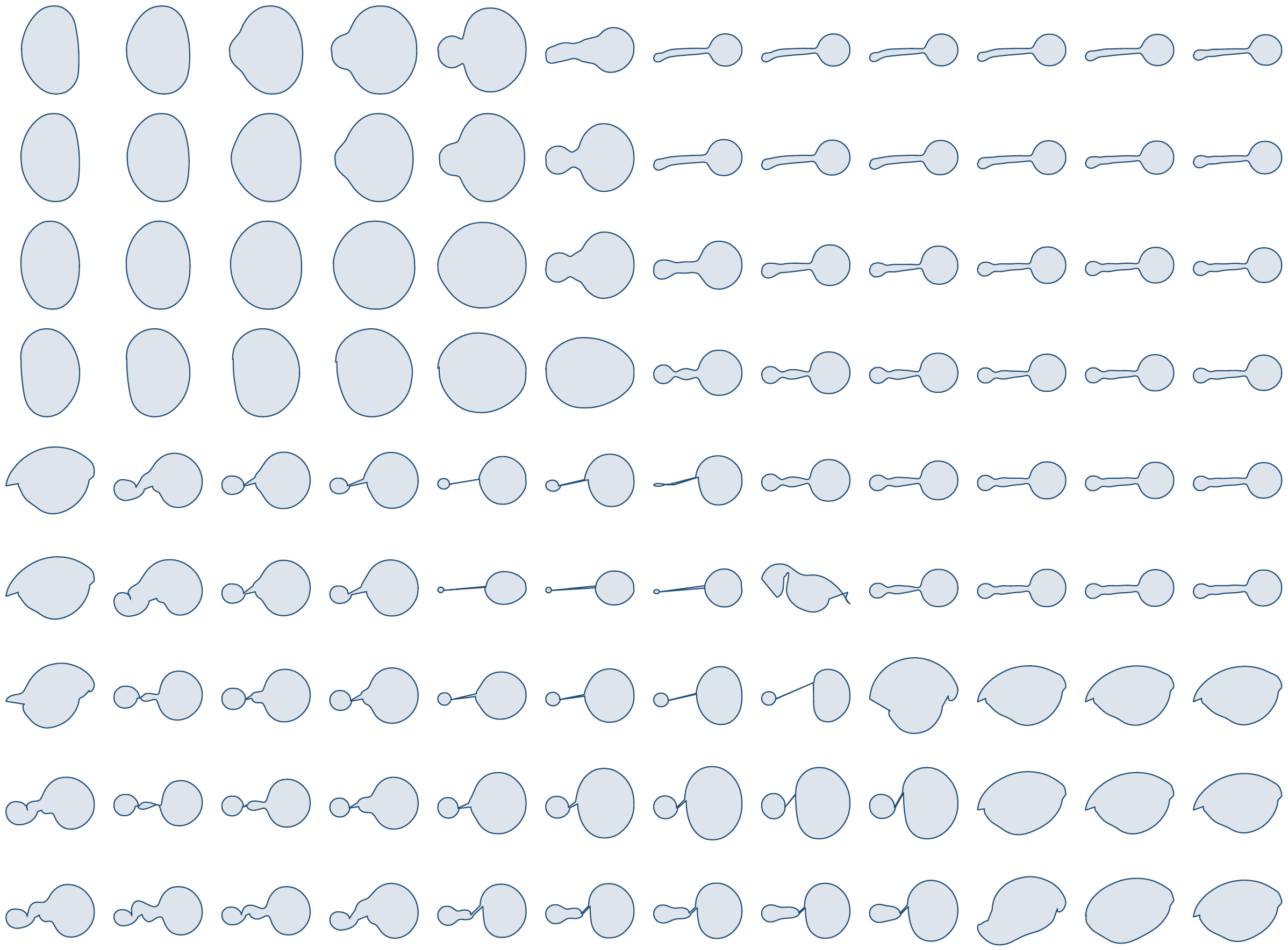}
    \caption{$\texttt{min\_dist}=0.8$ (continuous)}
  \end{subfigure}\hfill
  \begin{subfigure}{0.48\textwidth}
    \includegraphics[width=\textwidth]{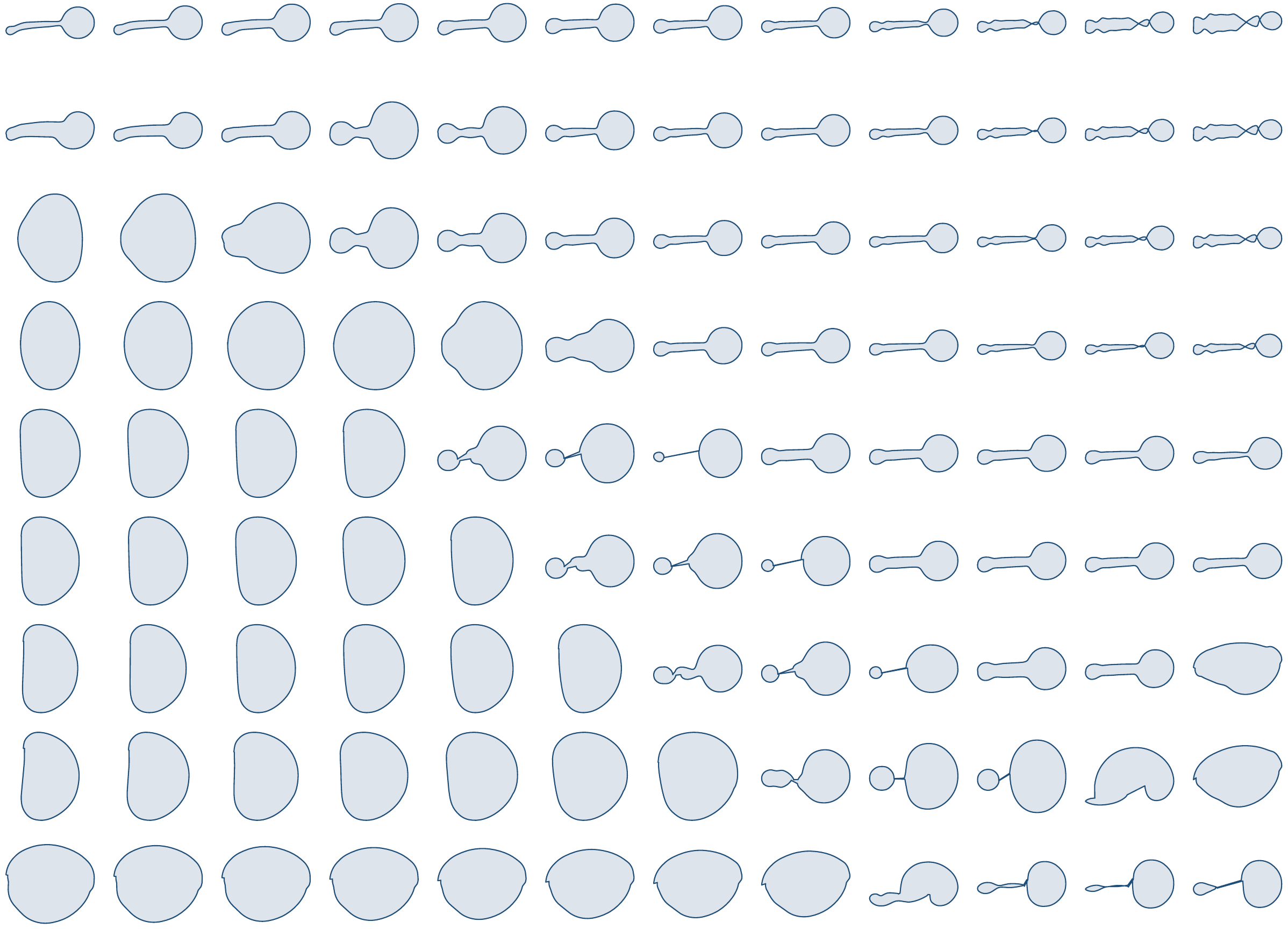}
    \caption{$\texttt{min\_dist}=0.01$ (manuscript)}
  \end{subfigure}
  \caption{Decoded latent grids under the two extreme regularisation
  strengths.}
  \label{fig:umap-grids}
\end{figure}

\subsection{Reproducibility}\label{sec:app:repro}
Dataset construction is deterministic from the raw frames given the recorded
configuration, which is stored as an attribute of the exported file together
with the rejection tally of \Cref{tab:cleaning}. Every reported value is the
mean over seeds $0$, $1$ and $2$, quoted with the sample standard deviation.
The raw contour dataset is public; see the data availability statement.

\typeout{get arXiv to do 4 passes: Label(s) may have changed. Rerun}
\end{document}